\documentclass[12pt,preprint]{aastex}
\usepackage{color}
\usepackage{lineno}
\graphicspath{ {./images/} }

\def\deg{\ifmmode {^\circ}\else {$^\circ$}\fi}
\def\degree{\ifmmode {^\circ}\else {$^\circ$}\fi}
\def\mum{\ifmmode {\rm \,\mu {\rm m}}\else $\rm \,\mu {\rm m}$\fi}
\def\arcsec{\ifmmode ^{\prime \prime}\else $^{\prime \prime}$\fi}

\def\inch{\ifmmode ^{\prime \prime}\else $^{\prime \prime}$\fi}
\def\msunyr{\ifmmode {M_{\odot}~{\rm yr^{-1}}}\else $M_{\odot}~{\rm yr^{-1}}$\fi}
\def\msun{\ifmmode {M_{\odot}}\else $M_{\odot}$\fi}
\def\rsun{\ifmmode {R_{\odot}}\else $R_{\odot}$\fi}
\def\lsun{\ifmmode {L_{\odot}}\else $L_{\odot}$\fi}
\def\mstar{\ifmmode {M_{\star}}\else $M_{\star}$\fi}
\def\rstar{\ifmmode {R_{\star}}\else $R_{\star}$\fi}
\def\tstar{\ifmmode {T_{\star}}\else $T_{\star}$\fi}
\def\lstar{\ifmmode {L_{\star}}\else $L_{\star}$\fi}
\def\md{\ifmmode {M_d}\else $M_d$\fi}
\def\ld{\ifmmode {L_d}\else $L_d$\fi}
\def\ad{\ifmmode A_d\else $A_d$\fi}
\def\ldlstar{\ifmmode L_d / L_\star\else $L_d / L_{\star}$\fi}
\def\rearth{\ifmmode {\rm R_{\oplus}}\else $\rm R_{\oplus}$\fi}
\def\mearth{\ifmmode {\rm M_{\oplus}}\else $\rm M_{\oplus}$\fi}
\def\qdstar{\ifmmode Q_D^\star\else $Q_D^\star$\fi}
\def\kms{\ifmmode {\rm km~s^{-1}}\else $\rm km~s^{-1}$\fi}
\def\ms{\ifmmode {\rm m~s^{-1}}\else $\rm m~s^{-1}$\fi}
\def\mesc{\ifmmode m_{esc}\else $m_{esc}$\fi}
\def\rmin{\ifmmode r_{min}\else $r_{min}$\fi}
\def\rmax{\ifmmode r_{max}\else $r_{max}$\fi}
\def\mmin{\ifmmode m_{min}\else $m_{min}$\fi}
\def\mmax{\ifmmode m_{max}\else $m_{max}$\fi}
\def\rmind{\ifmmode r_{min,d}\else $r_{min,d}$\fi}
\def\rmaxd{\ifmmode r_{max,d}\else $r_{max,d}$\fi}
\def\mmaxd{\ifmmode m_{max,d}\else $m_{max,d}$\fi}
\def\vrad{\ifmmode v_{rad}\else $v_{rad}$\fi}
\def\qz{\ifmmode q_{0}\else $q_{0}$\fi}
\def\qi{\ifmmode q_{i}\else $q_{i}$\fi}
\def\ql{\ifmmode q_{l}\else $q_{l}$\fi}
\def\qs{\ifmmode q_{s}\else $q_{s}$\fi}
\def\rbrk{\ifmmode r_{brk}\else $r_{brk}$\fi}
\def\rdamp{\ifmmode r_{damp}\else $r_{damp}$\fi}
\def\rin{\ifmmode r_{in}\else $r_{in}$\fi}
\def\rout{\ifmmode r_{out}\else $r_{out}$\fi}
\def\tin{\ifmmode t_{in}\else $t_{in}$\fi}
\def\tout{\ifmmode t_{out}\else $t_{out}$\fi}
\def\ain{\ifmmode a_{in}\else $a_{in}$\fi}
\def\aout{\ifmmode a_{out}\else $a_{out}$\fi}
\def\r0{\ifmmode R_{0}\else $R_{0}$\fi}
\def\m0{\ifmmode m_{0}\else $m_{0}$\fi}
\def\M0{\ifmmode M_{0}\else $M_{0}$\fi}
\def\xm{\ifmmode x_{m}\else $x_{m}$\fi}
\def\sigz{\ifmmode \Sigma_0\else $\Sigma_0$\fi}
\def\gyr{\ifmmode {\rm g~yr^{-1}}\else ${\rm g~yr^{-1}}$\fi}
\def\cms{\ifmmode {\rm cm~s^{-1}}\else ${\rm cm~s^{-1}}$\fi}
\def\gcms{\ifmmode {\rm g~cm^{-2}}\else $\rm g~cm^{-2}$\fi}
\def\gcmss{\ifmmode {\rm g~cm^{-2}~s^{-1}}\else $\rm g~cm^{-2}~s^{-1}$\fi}
\def\gcmc{\ifmmode {\rm g~cm^{-3}}\else $\rm g~cm^{-3}$\fi}
\def\dcm2{\ifmmode {\rm dyn~cm^{-2}}\else $\rm dyn~cm^{-2}$\fi}
\def\ecsk{\ifmmode {\rm erg~cm^{-1}~s^{-1}~K^{-1}}\else $\rm erg~cm^{-1}~s^{-1}~K^{-1}$\fi}
\def\cm2{\ifmmode {\rm cm^{-2}}\else $\rm cm^{-2}$\fi}
\def\atilin{\ifmmode {\tilde{a}_{in}}\else $\tilde{a}_{in}$\fi}
\def\atilout{\ifmmode {\tilde{a}_{out}}\else $\tilde{a}_{out}$\fi}
\def\atil{\ifmmode {\tilde{a}}\else $\tilde{a}$\fi}
\def\ttil{\ifmmode {\tilde{t}}\else $\tilde{t}$\fi}
\def\sqrttt{\ifmmode {\tilde{t}^{1/2}}\else $\tilde{t}^{1/2}$\fi}

\def\h2o{H$_2$O}
\def\sio2{SiO$_2$}
\def\ch4{CH$_4$}
\def\h2{H$_2$}
\def \ms{m\,s$^{-1}$\,}
\def \kms{km\,s$^{-1}$}
\def \msun{M$_{\odot}$}
\def \rsun{R$_{\odot}$}
\def \lsun{L$_{\odot}$}

\def \mearth{M$_{\oplus}~$}

\begin{document}
\shorttitle{Random Models}
\shortauthors{Podolak, Malamud, Podolak}
	
\title{Random Models for Exploring Planet Compositions I: Uranus as an Example}
\author{Joshua I. Podolak\altaffilmark{1}, Uri Malamud,\altaffilmark{2,3}, and Morris Podolak\altaffilmark{3}}

\altaffiltext{1}{D. E. Shaw Group, New York, NY}
	\altaffiltext{2}{Department of Physics, Technion - Israel Institute of Technology, Technion City, 3200003 Haifa, Israel}
	\altaffiltext{3}{School of the Environment and Earth Sciences, Tel Aviv University, Ramat Aviv, 6997801 Tel Aviv, Israel}

\begin{abstract}
Modeling the interior of a planet is difficult because the small number of measured parameters is insufficient to constrain the many variables involved in describing the interior structure and composition.  One solution is to invoke additional constraints based on arguments about how the planet formed.  However, a planet’s actual structure and composition may hold clues to its formation which would be lost if this structure were not allowed by the initial assumptions.  It is therefore interesting to explore the space of allowable compositions and structures in order to better understand which cosmogonic constraints are absolutely necessary.  To this end, we describe a code for generating random, monotonic, density distributions, $\rho(r)$, that fit a given mass, radius, and moment of inertia.  Integrating the equation of hydrostatic equilibrium gives the pressure, $P(r)$, at each point in the body.  We then provide three algorithms for generating a monotonic temperature distribution, $T(r)$, and an associated composition that is consistent with the $\rho-P$ relation, and realistic equations of state.  We apply this code to Uranus as a proof of concept, and show that the ratio of rock to water cannot be much larger than 2.
\end{abstract}
	
\section{Introduction}\label{S:Intro}
With the discovery of nearly 5000 exoplanets with densities ranging from less than 0.03 \citep{masuda2014} to more than 10 times Jupiter's density \citep{marcy2014}, it becomes a challenge to understand the composition of these bodies. Even within our own solar system, where the parameters characterizing the different planets are well known, it can still be difficult to determine a definite composition.  An example of this difficulty is Jupiter, where we have not only accurate measurements of the mass and radius, but also of the first few gravitational moments \citep{iess2018} yet there is still an ambiguity regarding the total mass of high-Z material \citep{wahl2017, debras2019}.  The simple two- and three-layer models of the past which assumed adiabatic interiors for the outer planets \citep{podcam74,fortnet10,nettel12b} are being replaced by more sophisticated models which allow for a continuously changing interior composition, and non-adiabatic thermal profiles \citep{leconte12,helstev2017,vazan2020,vazanhel2020}.

Because the internal thermal, chemical, and physical structure of a planet can be complex, attempts have been made to glean information about the possible composition of the outer planets by side-stepping the usual physical models, and looking for internal density distributions, $\rho(r)$, that would fit the observed parameters, and could then be interpreted in terms of composition. \cite{helled08b} and \cite{helledetal10} took $\rho(r)$ to be a polynomial, while \cite{movshovitz2020} used a combination of polynomials.  \cite{marley95} and \cite{podpod00} went further, and took $\rho(r)$ to be a random monotonic function.  Although this allowed for more freedom in the shape of $\rho(r)$, it required assumptions about the corresponding temperature profile, $T(r)$, in order to say something about the composition.  In what follows, we return to these random distributions, but with an improved algorithm that allows us to efficiently sample a broad range of density distributions as well as to make some arguments about possible compositions.

Our intention is to demonstrate that such random models can, indeed, place some constraints on possible compositions of planets in general, and for this study we focus on Uranus in particular. We view this work as a ``proof of concept" for this idea, and in a series of papers plan to extend it to other applications using the approaches and tools developed here.

Further motivation for our work arises from our feeling that giant planet composition must be examined more rigorously, and without relying on a preconceived conceptual framework (bias?) that tends to regard the outer giants of our solar system  as 'icy' (i.e. water-rich). Uranus, the focus of our current investigation, and Neptune have formed far from the Sun, where water condensates would seem to be abundant building blocks. However, there is no conclusive evidence that these initial constituent building blocks are in fact water-rich. Indeed, there are several arguments to indicate that the opposite might be true. 

First, there is evidence that solids in the adjacent Kuiper belt region are rock-rich. For example, \cite{malamud2015} have utilized thermophysical evolution models of various Kuiper belt dwarf planets to show that the known trend in Kuiper belt objects of increasing bulk density with mass \citep{brown2012} can arise solely from differences in their internal porosity, due to thermal and mechanical processing.  It is assumed that since Kuiper belt objects all form in the same region, they also have a similar basic composition  (see e.g. \cite{Kenyon2008}). \cite{malamud2015} found that a good match for the composition is a rock to water mass ratio of $\sim$3, which fits the observed trend. Later work by \cite{bierson2019} arrives at the same conclusion.  A large rock content is also directly inferred from the bulk density of the largest Kuiper belt dwarf planets, such as Pluto.  Pluto has very little residual porosity  \citep[see, e.g.][]{bierson2018}, and its measured bulk density explicitly translates to rock-rich material. Other large Kuiper belt dwarf planets sometimes show evidence of even higher bulk densities, typically associated with additional modifications through giant collisions which preferentially remove water from an existing differentiated structure (see e.g. \cite{lissauer1993} and \cite{barr2016} for Eris, or \cite{leinhardt2010} for Haumea).

Second, modern theories for comets, which originate in the outer solar system, and which represent some of its most pristine and basic building blocks, regard comets as highly rock-rich bodies. There is a growing body of evidence that strongly suggests comets have rock to water mass ratios of \emph{at least} 3, and potentially even twice that value \citep{rotundi15,fulle2016,fulle2017,fulle2019,choukroun2020}.  Likewise, the dust-to-water-ice mass ratio measured in molecular clouds is close to 5 \citep{cambianica2020}.

Finally, exoplanet discoveries also place interesting constraints for high rock content outside the solar system, according to some widely favored models. We refer to the observed 'valley' in the radius distribution of small exoplanets, in which exoplanets in the range 1.5-2 R$_\oplus$ are significantly less common than smaller or larger exoplanets. The two main channels to explain this bimodality are: loss of atmosphere by stellar photons \citep{owen2013} (the photoevaporation channel), and intrinsic luminosity of the cooling planet cores, which can erode light envelopes while preserving heavy ones \citep{ginzburg2018} (the core-powered mass-loss channel). Both channels produce a deficit of intermediate sized planets and at present either one is equally plausible from an observational standpoint \citep{rogers2021}. According to core-powered mass-loss, however, models that fit the observations remarkably well, also specifically require rock-rich cores. In contrast, model outcomes with fully icy or iron cores do not resemble the observations, and produce a significant shift from the required planet sizes (see Figure 4 in \cite{gupta2019}). This latter channel would imply that (a) close-in exoplanets could have formed much further out in the disk than their current orbit suggest; and (b) Uranus and Neptune could be similar to these sub-giant exolplanets and perhaps be composed of rock-rich cores.

In conclusion, while a high rock to water mass ratio is suggested for the nearby Kuiper belt, for comets of similar origin, and perhaps even for exoplanets, the current view is that for the outer giants the reverse is true in terms of their composition.  They are viewed as having a rock to water ratio of roughly 0.5, consistent with solar composition.  This calls for a second look at the fundamental issue of sub-giant planet composition. At the very least we should make an effort to quantify the likelihood, or the conditions, required for producing successful rock-rich models that match the observations.  Here we explore the possibility of such rock-rich models for Uranus.

The paper outline is as follows: in Section \ref{S:Input} we discuss the input data to which our random models are compared; in Section \ref{S:Random} we present the basic ideas behind our algorithms for computing random models; in Section \ref{S:Matching} we show how to generate random models that also match our input data criteria; in section \ref{S:TemperatureComposion} we explain our strategy for how to associate an appropriate temperature and composition profile with a given density profile.  In Section \ref{S:Results} we present some tentative results, and discuss directions for future work, and we summarize our conclusions in Section \ref{S:Summary}.

\section{Input Data}\label{S:Input}
For exoplanets, in general, only the mass and radius can be measured with any accuracy.  For Uranus, in addition, the gravitational moments, $J_2$ and $J_4$ \citep{jacobson2014} have been measured.  However, in order to compute these moments for a given $\rho(r)$, the rotation period of the body must be known, and for Uranus there is some ambiguity in this matter \citep{heletal10}. In addition, the computation of the moments is time-consuming, so we have chosen, instead, to fit the moment of inertia.  While the moment of inertia has not been directly measured, models of Uranus that do fit the observed $J_n$'s give a normalized moment of inertia (MOI) $I/MR^2\sim0.23$ \citep{heletal10,nettel12b} and this is the value we have chosen to match.  Thus our goal is to explore the space of possible functions $\rho(r)$ that fit Uranus' measured mass and radius, as well as an assumed MOI = 0.23.  

Another constraint is that $\rho(r)$ has to be monotonically increasing from the surface of the planet to its center.  Furthermore, since we are interested in relating $\rho(r)$ to possible compositions, we impose the additional constraint that if $X_i$ is the mass fraction of the $i$th component, the different $X_i$'s must vary in such a way that the average mass fraction of elements heavier than helium (hereafter called \textit{metals}), $Z$, increases monotonically from the surface to the center.  While it is true that small planetary bodies can have stratified structures leading to non-monotonic mass fractions as a function of radius \citep[see, e.g.][]{malamud2017}, larger bodies, such as Uranus, are expected to have strong enough mixing so that the monotonic mass fraction assumption is justified.  Our task, then, is to generate random distributions of density as a function of radius that satisfy the above constraints, and then to determine what these can tell us about possible limits on the composition of Uranus.

\section{Generating Random Models}\label{S:Random}
As we have noted above, we not only want to generate random 1-D functions of density as a function of radius, but, we also want to have a random, monotonic variation in the average value of the metal mass fraction.  Our goal in this section and the next is to describe a method for generating the density models for a planet.  The computation of the composition and temperature profiles is described in Section \ref{S:TemperatureComposion}.  Our model consists of a fixed number of shells, each with a density value.  For now, the only constraints on the models are that it match a given Mass/Radius/MOI, and that the densities are monotonic as a function of radius.

The first observation we make is that the for every valid density model, we could simply divide by the maximum density and obtain a monotonic function, $\bar{\rho}$, with a maximum value of one.   We can similarly normalize the max radius to one.   
Thus, we have reduced the problem of finding valid density models to finding monotonic functions $f:[0,1]\rightarrow[0,1]$

Our second observation is that we want our models to be varied.   That is, we want to generate models that look like straight lines, various power law curves, functions with large jumps in density, etc.   More generally stated, we want our models to ``cover" the space of \emph{all} possible monotonic functions $f:[0,1]\rightarrow[0,1]$.   The challenge is to come up with a method that has a reasonable chance of covering all possibilities.

\subsection{Sorting}
To understand the challenge, consider what is possibly the easiest way to come up with a random monotonic function.   Assuming we require a model with $K$ shells, we could simply generate $K$ uniformly-distributed numbers in the range [0,1], sort them, and assign them sequentially to the shells.   Models generated will be completely random, guaranteed to be monotonic, and due to the Law of Large Numbers, \emph{converge to the diagonal y=x as the number of shells increases}.   For an example of this law, consider the likelihood of getting a model with a density jump from 0.1 to 0.2.  Assuming there are $K$ shells, this would require $K$ independently generated random numbers to all be out of the range [0.1,0.2], giving us a probability of $P=0.9^K$.  For $K=100$, the probability evaluates to $2.5\times 10^{-5}$, and decreases exponentially with the number of shells. 

Note that convergence happens regardless of the distribution we choose values from; it just converges to a different curve (it can be shown that convergence is related to the inverse of the cumulative distribution function of the distribution).   One possible solution would be to have a set of varied distributions to draw points from, but it would be hard to choose a sufficiently varied set so that all possible models are reasonably likely to occur.

\subsection{Early Random Model Attempts}
\cite{marley95} and \cite{podpod00} tried to generate random monotonic density functions by starting at the outermost mass shell with some random, small, density and taking the density of successive inner shells to be some random value between the density of the previous shell and the mean density of the remaining shells. This is equivalent to choosing values in the range $[0,1]$ and multiplying by the factor appropriate to the mass remaining. This method tends to reach the maximum value quickly and remain constant afterwards. Podolak et al. (2000)  attempted to mitigate this problem by sampling values from a uniform distribution and then raising those values to the power of a fixed constant $C$.   (Optimally, $C$ is chosen as a function of the number of shells $K$, so that the expected number of shells before hitting maximum value is $\beta K$ where $0.5\leq \beta \leq 1$).  Their method did find models outside the previous range, but only with low probability.

\subsection{Hierarchical Methods}

The best way to avoid convergence is to assign most of the variance to a single random number.   Consider the following example:   Assume that we wished to generate a uniformly-distributed random number in the range $[0,1]$, but all we had was a fair coin.   The solution would be to write the random number as a binary fraction and ``flip" the coin for each bit (up to the level of accuracy needed).   One reason this avoids the Law of Large Numbers is because most of the variance is in the first coin flip.

Similarly, if we first fix the gross features of the model with very few random values, we can statistically cover the entire spread of possibilities.   We can then choose additional random values to fill in the details.   At the extreme, we would start with two shells and use a single random number $0<\bar{\rho}_1<1$.   The outer shell would have a density in the range $[0,\bar{\rho}_1]$, and the inner shell would have a density $[\bar{\rho}_1,1]$.   Then, at every step, we subdivide the shells, randomly choosing appropriate densities that keep the entire model monotonic.  This method avoids convergence, but does have the weakness that most of the variance comes from choosing the density of the middle shell (at $r=R/2$).   

\begin{figure}[h!]
	\centering
	\includegraphics[height=12cm]{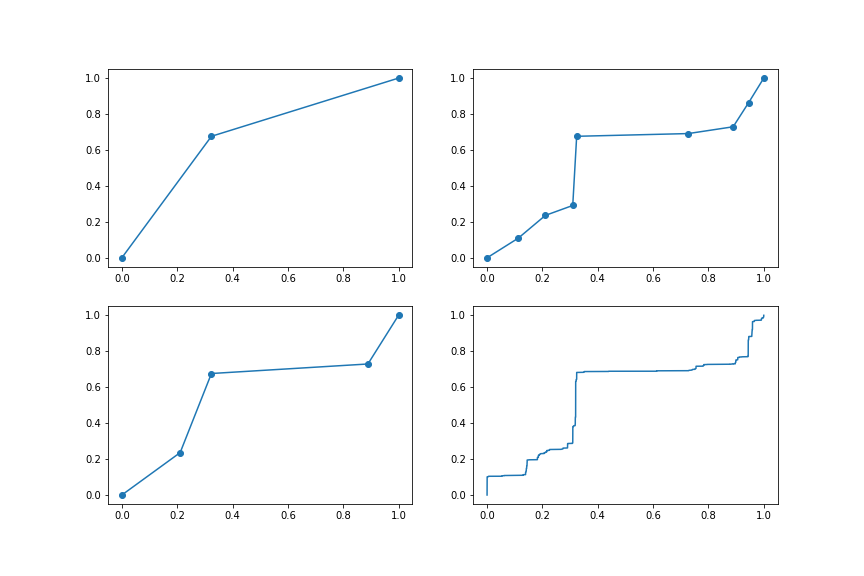}
	\caption{{\bf Upper left:} One iteration of MSP method. 
		{\bf Lower left:} Two iterations.
		{\bf Upper right:} Three iterations.
		{\bf Lower right:} Ten iterations}
	\label{fig:raster_recursive_diagonal}
\end{figure}

We can correct this weakness by relaxing the constraint that we always choose the initial two shells to be equal in size.  Instead, we will randomly choose where the shell-breaks are.   Our final algorithm is as follows:

\begin{verbatim}

Given K uninitialized shells:
    Randomly choose i = random{1,,K} and set shell[i] = random(0,1).
    Keep randomly selecting uninitialized shells:
        For each shell, randomly choose a density such that monotonicity is not broken.

\end{verbatim}
An illustration of this \textit{most significant point} (MSP) method is shown in Fig.\,\ref{fig:raster_recursive_diagonal}.

\section{Matching Random Models with Input Data}\label{S:Matching}
\begin{figure}[h!]
    \centering
	\includegraphics[height=4.5cm]{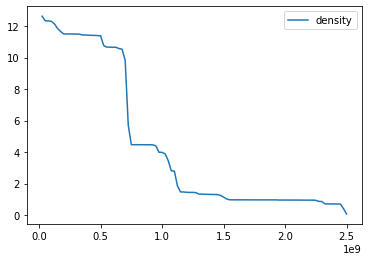}
	\includegraphics[height=4.5cm]{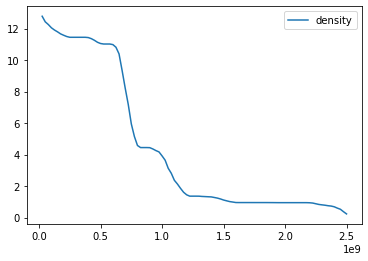}
	\includegraphics[height=4.5cm]{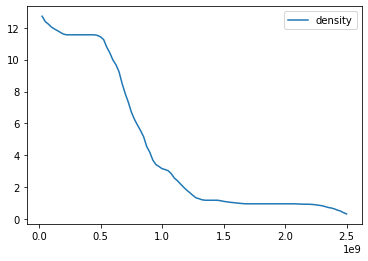}
	\includegraphics[height=4.5cm]{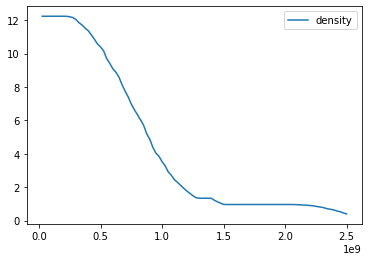}
	\caption{ The effects of smoothing on a random model using a Savitzky-Golay filter \citep{savitzky1964}.  
	         {\bf Upper left:} No smoothing. 
	         {\bf Upper right:} kernel size = 100. 
	         {\bf Lower left:} kernel size = 200.
	         {\bf Lower right:} kernel size = 400.}
	\label{fig:smoothing_example}
\end{figure}

We use the above algorithm to generate planetary models that match a given mass, radius and moment of inertia. We do this by splitting the planet into a series of concentric shells and randomly choosing a monotonically increasing density for each shell.  Note that our algorithm constructs models where the innermost density is lowest so we need to reverse the order of the shells.   In addition, we find that the models created by our algorithm tend to contain many small abrupt jumps, something we do not expect for the density distributions in planetary interiors.  We therefore used an off-the-shelf implementation of the Savitzky-Golay filter \citep{savitzky1964} to smooth the models.   Fig.\,\ref{fig:smoothing_example} shows the effects of smoothing with different sized kernels.

The mass of a planet may be found by taking the integral $ 4\pi\int r^2\rho dr $.   The key point here is that the mass is a linear function of density.  This means that if we find that our initial model has a mass of $M$ and we multiply all the densities in the model by a constant $\alpha$, the new model will have a mass of $\alpha M$.  Multiplying all the density values by a fixed amount does not break monotonicity, thus it is very easy to obtain a model that matches a given mass.

Similarly, if we have two models, $C_1$ and $C_2$, that both have mass $M$, then any linear combination of them $C' = \alpha C_1 + (1-\alpha) C_2$ will also have mass $M$.   As long as the two initial models $C_1$ and $C_2$ are monotonic, and $0<\alpha<1$, then the blended model will also be monotonic.  

The MOI of a planet may be found by taking the integral $ \frac{8}{3}\pi\int r^4\rho dr $.  As with mass, the MOI is a linear function of density.  Our method for matching both mass and moment is to create a large number of random density models (100 by default) that have all been normalized to have the correct mass.   We then compute the MOI for each model and split them into two sets: those with an MOI greater than the desired value and those with an MOI smaller than the desired value.  The large number of models we generate ensures that both sets are not empty.  We randomly choose one model from each set.  Both models have the same mass, so any positive linear combination of the two models will match the desired mass and maintain monotonicity.   As noted above, the MOI is a linear function of density, so it is easy to find a linear combination that matches the desired value.

It is worth noting that any positive linear combination of two models that match both mass and MOI will also match the mass and MOI.  Thus, one could potentially match a further global parameter such as the gravitational moment, $J_2$.  Although the gravitational moments are not a linear function of density, there is only one free variable, so a method like gradient descent or binary search should suffice.   Obtaining models that match $J_2$ in addition to mass and moment is a potential goal for future work.
 
\section{Associated Temperature and Composition}\label{S:TemperatureComposion}
Once a density profile $\rho(r)$ matching the radius, mass and MOI of a particular planet has been generated, there is still a large degree of freedom in how one chooses the associated temperature and composition profiles. Until recently, models of giant planets have assumed a 2- or 3-layer structure, composed of single, unmixed materials which were typically rock, water and a gaseous H-He envelope \citep{podcam74,fortnet10,nettel12b}. Such models are nowadays being modified according to the assumption that during planetary growth the accreting heavy elements can mix, resulting in internal compositional gradients \citep{leconte12,helstev2017,vazan2020}. The details of mixing are different in various studies. Some models permit mixing of rock and envelope materials \citep{helledetal10,vazanhel2020}, while others permit mixing of either just the heavy elements, materials from adjacent layers \cite{vazanhel2020}, or mixing of all three materials \citep{vazan2020}.

In what follows, we assume four different materials: a solar mix of 72.5\% hydrogen and 27.5\% helium by mass, which we refer to as {\it envelope}, water, \sio2, and iron.  We allow for mixing only between neighboring materials, thus only the following mixtures are considered: envelope + water, water + \sio2, and \sio2 + iron.  In these mixtures the mass fraction of the heavier material increases monotonically towards the center.  By including the possibility of pure iron, we allow for potentially higher central densities when selecting random density models. For the iron, \sio2 and water we use the quotidian equation of state \citep{more88, vazan13} and for envelope material we use SCvH equation of state \citep{saumchab95} as modified in \citep{chabrier2019}. For mixing of different materials we use the additive volume law.

We then proceed as follows:  First, we generate a random, monotonic function $\rho(r)$ that satisfies the mass, radius, and MOI as described above.  Given the mass distribution, $M(r)$, we can integrate the equation of hydrostatic equilibrium
\begin{equation}
	\frac{dP}{dr}=-\frac{GM(r)\rho(r)}{r^2}
\end{equation} 
to find the pressure, $P(r)$.  Here $G$ is Newton's constant.  Given the pressure and density at every point in the planet, one can then assign a composition and determine the temperature, $T$, required so that the composition gives the desired density at the given pressure.  In order to limit the infinitely many combinations of composition and $T$ that will satisfy the $P(\rho)$ curve we have generated, we impose the additional physical restrictions that both the composition and $T$ vary monotonically with $r$ as described above.  This, of course, still leaves us with infinitely many possibilities, but in the spirit of looking for models that allow a composition similar to KBO's, as discussed in detail in Section \ref{S:Intro}, we conduct our search with the goal of finding compositions that have large overall rock to water ratios. In the following sections we present three algorithms for producing such monotonic variations of composition and temperature.

\subsection{The Inward Algorithm}\label{SS:inverse}
In this algorithm we start by matching the temperature/composition of the random density model at the planet's outer boundary, and move inwards towards the center.  We first have to assume an appropriate temperature for the surface, $T_s$ (in Uranus' case we choose 70\,K at 1\,bar, as in \cite{podpod00}). With $T=T_s$, we find an initial composition of either pure envelope or a mixture of envelope + water that matches the desired surface density and pressure. We continue inwards, and at each consecutive shell we attempt to minimize the mass fraction of high-Z material whenever possible (first water, then rock, then iron) while keeping the temperature the same. If there is no mix that matches the corresponding density and pressure, we increase the temperature while keeping the same composition.  This procedure maximizes the low-Z material content and keeps the temperature as low as possible.

This {\it inward algorithm} is always successful, unless the central density in the random model is so high that even a pure iron composition is unable to match it. Such models are considered invalid and are discarded.  In practice, we find that such high central densities occur only infrequently.  

\begin{figure}[h!]
	\centering
	\includegraphics[height=5.5cm]{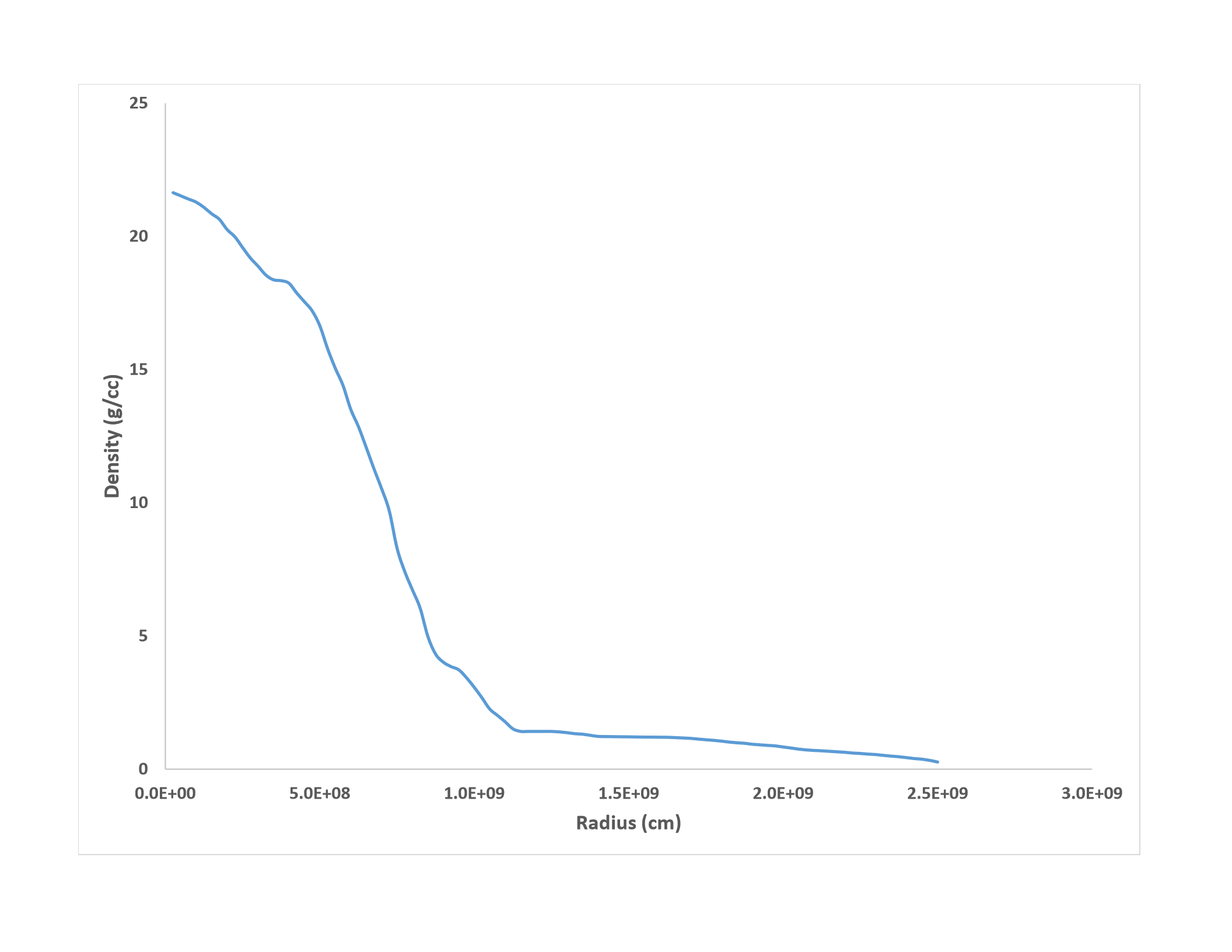}
	\hspace{1cm}
	\includegraphics[height=5.5cm]{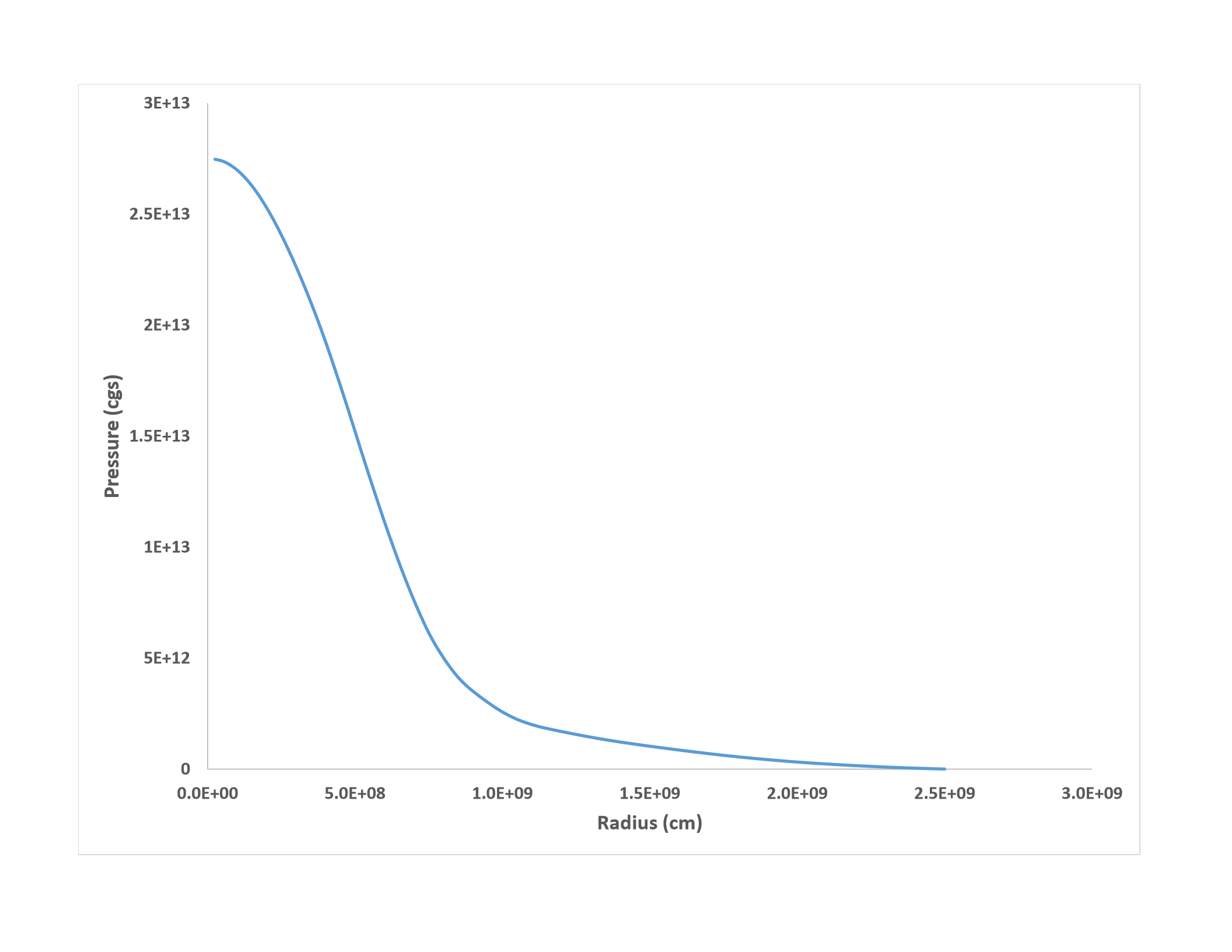}
	\caption{An example of a randomly generated model.  {\bf Left:} Density as a function of radius.  {\bf Right:} Pressure as a function of radius.}
	\label{fig:random_model_example}
\end{figure}

\begin{figure}[h!]
	\centering
	\includegraphics[height=5cm]{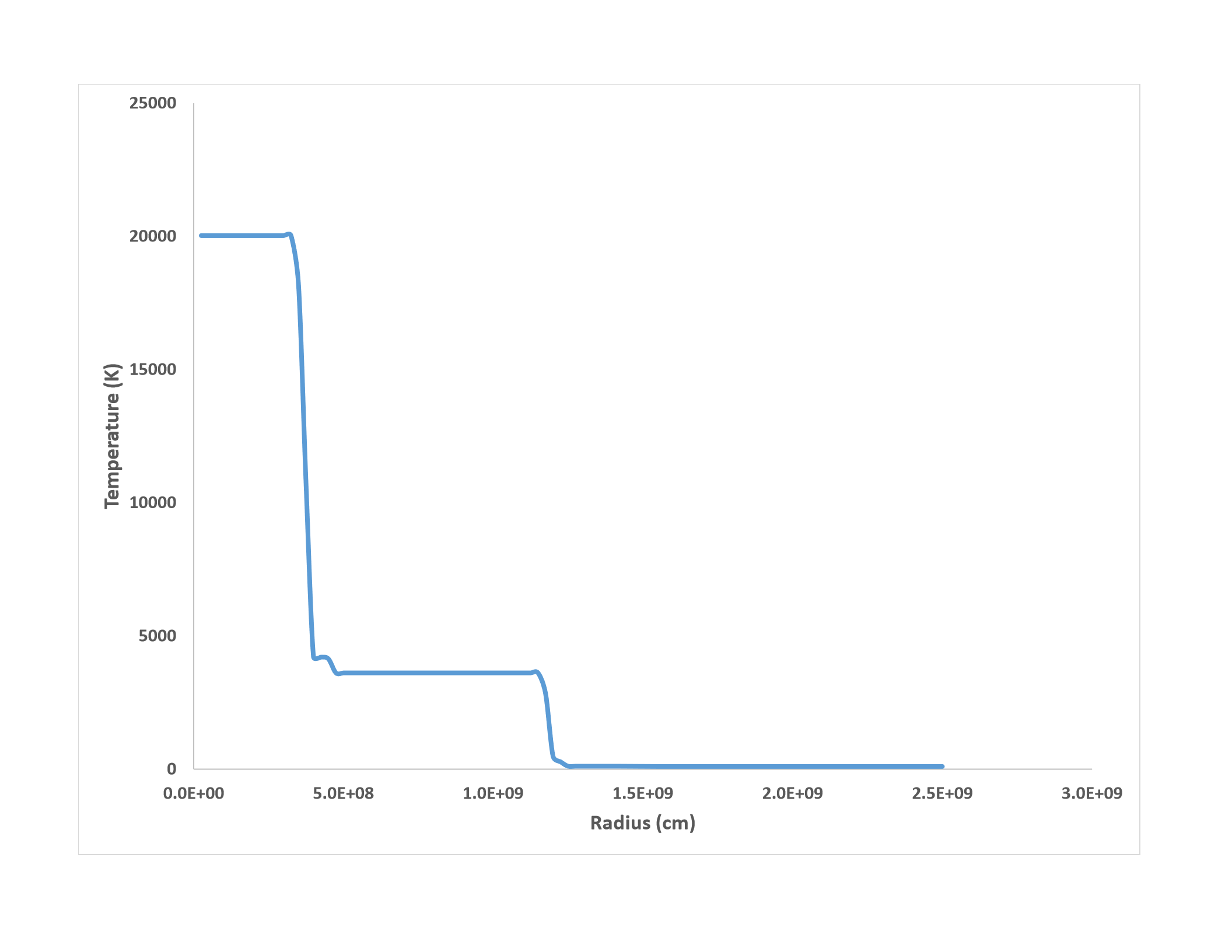}
	\hspace{1cm}
	\includegraphics[height=5cm]{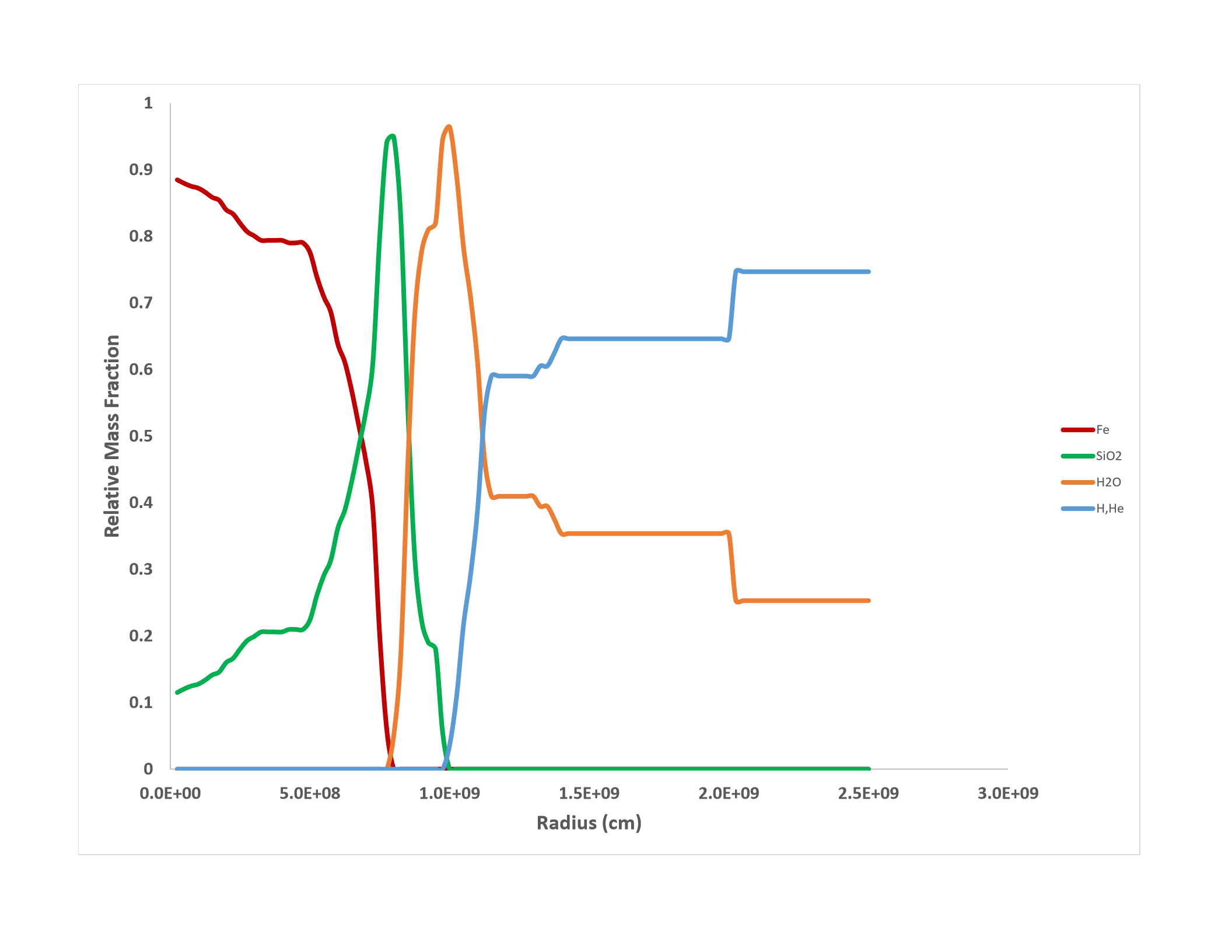}
	\caption{The derived temperature and composition profile corresponding to the model in Fig.\,\ref{fig:random_model_example}, using the inward algorithm. {\bf Left:} Temperature as a function of radius. {\bf Right:} Mass fraction of Fe (red), \sio2 (green), water (orange) and envelope (blue) as a function of radius.}
	\label{fig:inverse}
\end{figure}

As an example, consider the model in Fig.\,\ref{fig:random_model_example}, which shows a randomly generated density profile (left panel) and the corresponding pressure profile (right panel). 
 Fig.\,\ref{fig:inverse} shows the corresponding temperature profile (left panel) derived using the inward algorithm. The composition is shown in the right-hand panel.  Starting from the outer edge of the planet the composition is a mixture of roughly 75\% envelope and 25\% water.  As one moves inward, the water mass fraction increases to nearly 1 and the envelope mass fraction decreases to nearly 0 at around $r=1.0\times 10^9$\,cm.  The \sio2 and iron mass fractions are 0 throughout this region.  After that the \sio2 mass fraction starts to increase and the water mass fraction decreases.  The latter reaches 0 at $r=8\times 10^8$\,cm.  At still smaller radii even pure \sio2 cannot match the $P(\rho)$ relation, and we must add iron.  The central composition is approximately 12\% \sio2 and 88\% iron. The central temperature attained for this particular model is around $T_c=2\times 10^4$\,K. This algorithm allows water in the outer regions only ``when necessary", and in this way tries to maximize the rock to water ratio in this region.  It also maximizes the envelope mass fraction.  For this particular case the rock to water ratio for the planet is 1.2.

\subsection{The Outward Algorithm}\label{SS:forward}
A second possible algorithm would be to start matching the random density model at the planet's center and move outwards towards the surface. Unlike the case for the inward algorithm, where the surface temperature can be determined, here we must start from some postulated central temperature, $T_c$. Using  $T_c$ as our constraint, we find an initial composition of either our heaviest material or a mixture of two of our heaviest materials, that matches the desired central density and pressure.  We continue outwards, and at each successive shell we check to see if the same composition can match the density and pressure at that point, while keeping the temperature no higher than that of the previous shell (i.e. monotonically decreasing). Otherwise, we find a composition that retains the previous temperature by decreasing the metal content (first iron, then rock, then water).

This outward algorithm ``prefers" lower water content in the inner mass shells and in this way tries to maximize the rock to water ratio.  However, since the central temperature is an initial guess, there are occasionally cases where we cannot find a suitable composition in the equation of state look-up tables.  In such cases we can simply ignore the shell and skip to the next. If a given random density profile does not produce many such null points, the problem can generally be overcome by applying extra smoothing to the density profile. Besides the necessity of guessing a central temperature and the occasional rare null points, another potential problem of the outward algorithm is that it can often generate planet compositions lacking any envelope, or having merely a negligible envelope. This is not necessarily a problem for Super-Earths, but such results are obviously not appropriate for giant or sub-giant planets.

\begin{figure}[h!]
	\centering
	\includegraphics[height=5cm]{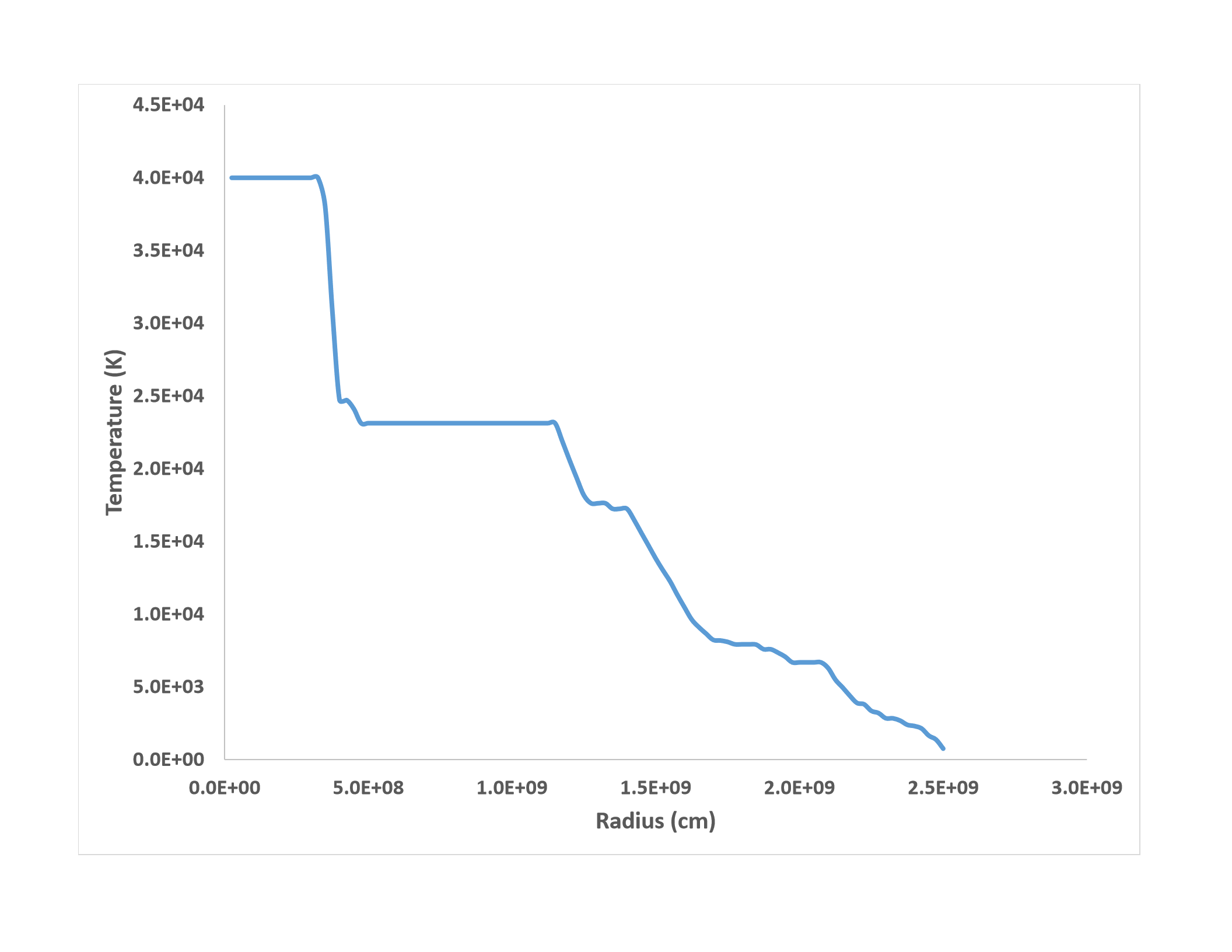}
	\hspace{1cm}
	\includegraphics[height=5cm]{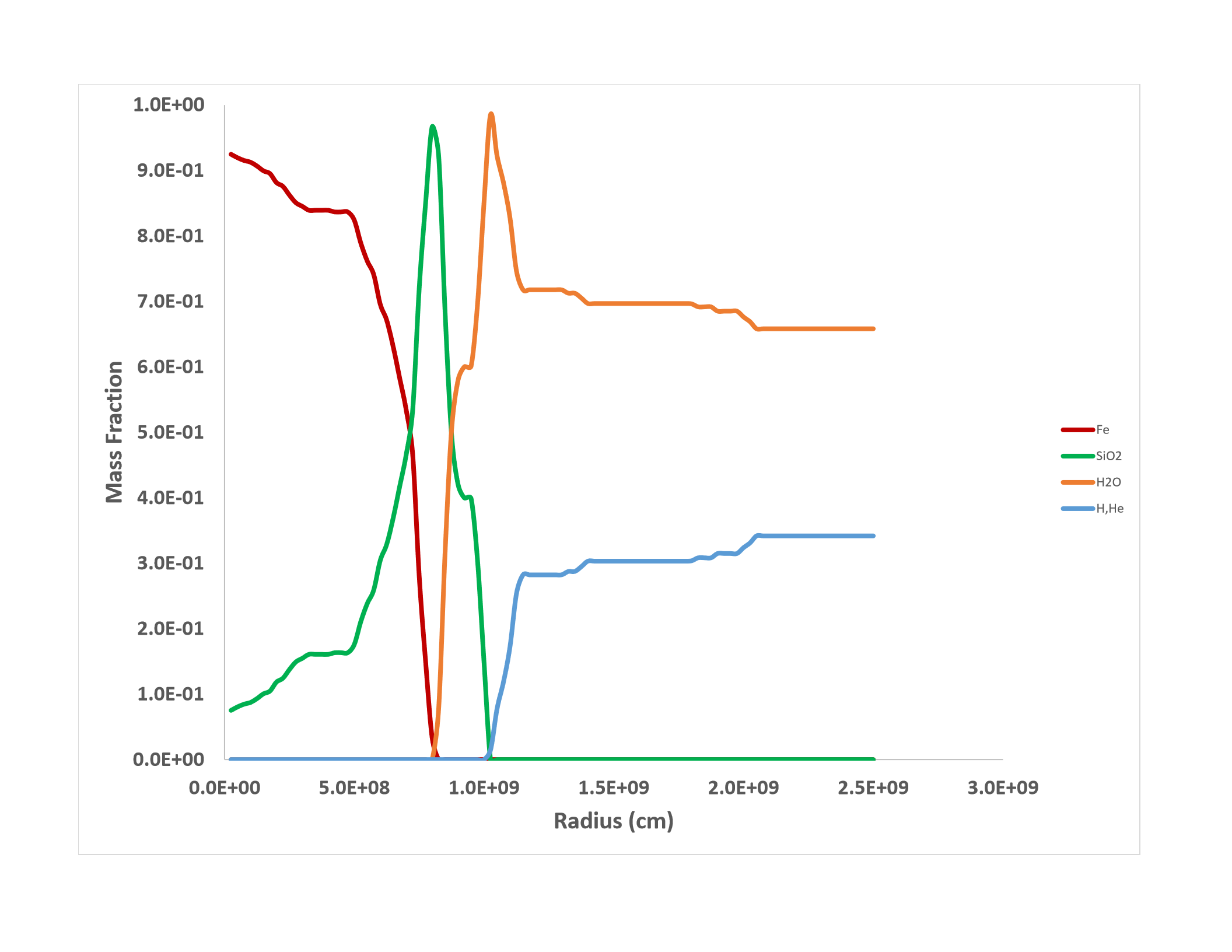}
	\caption{The derived temperature and composition profile corresponding to the model in Fig.\,\ref{fig:random_model_example}, using the outward algorithm. {\bf Left:} Temperature as a function of radius. {\bf Right:} Mass fraction of Fe (red), \sio2 (green), water (orange) and envelope (blue) as a function of radius.}
	\label{fig:forward}
\end{figure}

As expected, when we run the outward algorithm with the same central temperature as derived from the inward algorithm, $T_c=2\times 10^4$\,K, we get a model that is nearly identical to that of the inward algorithm.  To gauge the effect of the central temperature, we ran the same model with $T_c=4\times 10^4$\,K.  The result is shown in Fig.\,\ref{fig:forward}.  The composition that now emerges has somewhat more iron near the center, and the total rock mass increases from 5.1 to 5.3\,$M_{\oplus}$. However the water mass increases even more, from 4.1 to 6.9\,$M_{\oplus}$.  As a result, the rock to water ratio decreases to 0.77.  The outer layers are also more water-rich, with water comprising 66\% by mass of the outermost layer, as opposed to 25\% for the inward algorithm.

\subsection{The Bi-directional Algorithm}\label{SS:bi-directional}
\begin{figure}[h!]
	\centering
	\includegraphics[height=5cm]{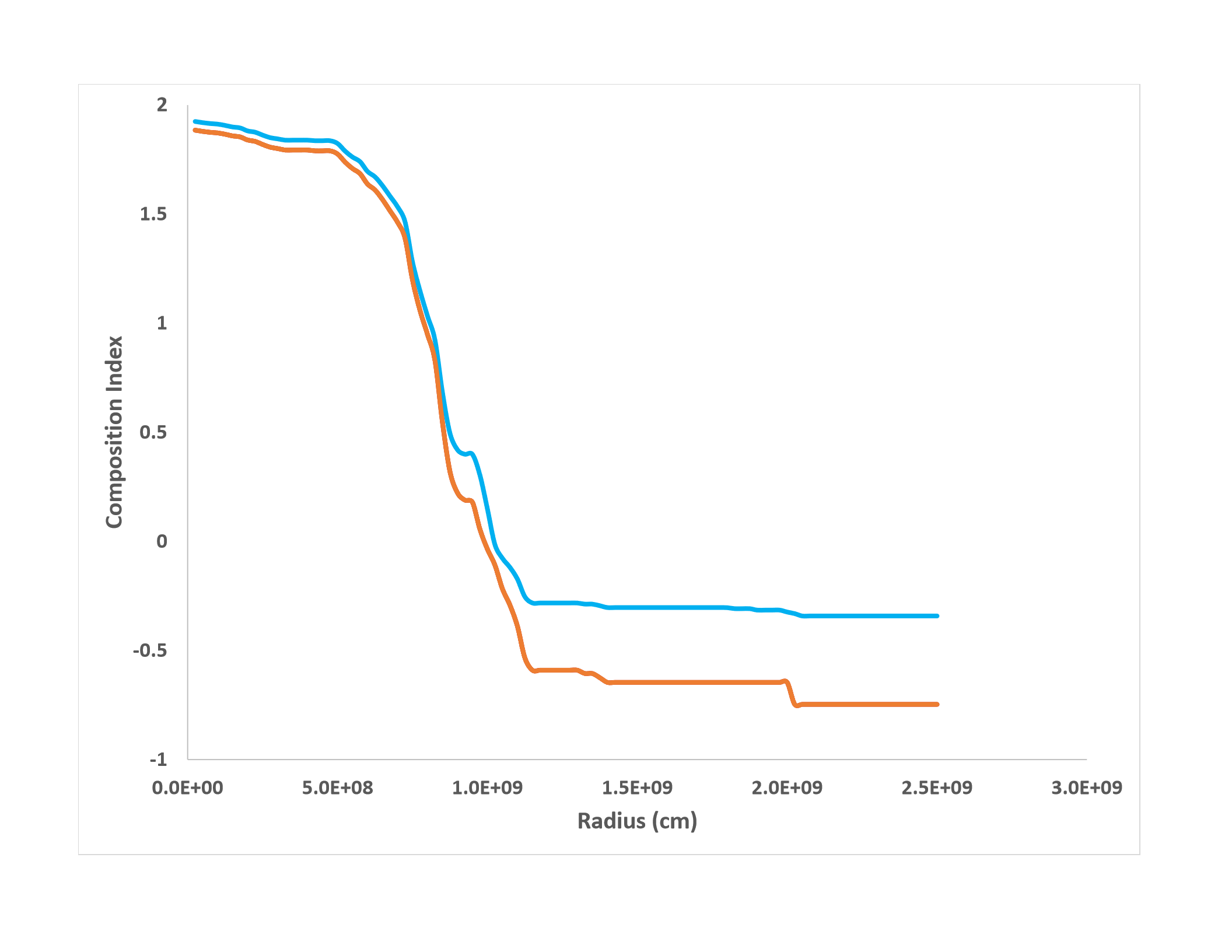}
	\hspace{0cm}
	\includegraphics[height=5cm]{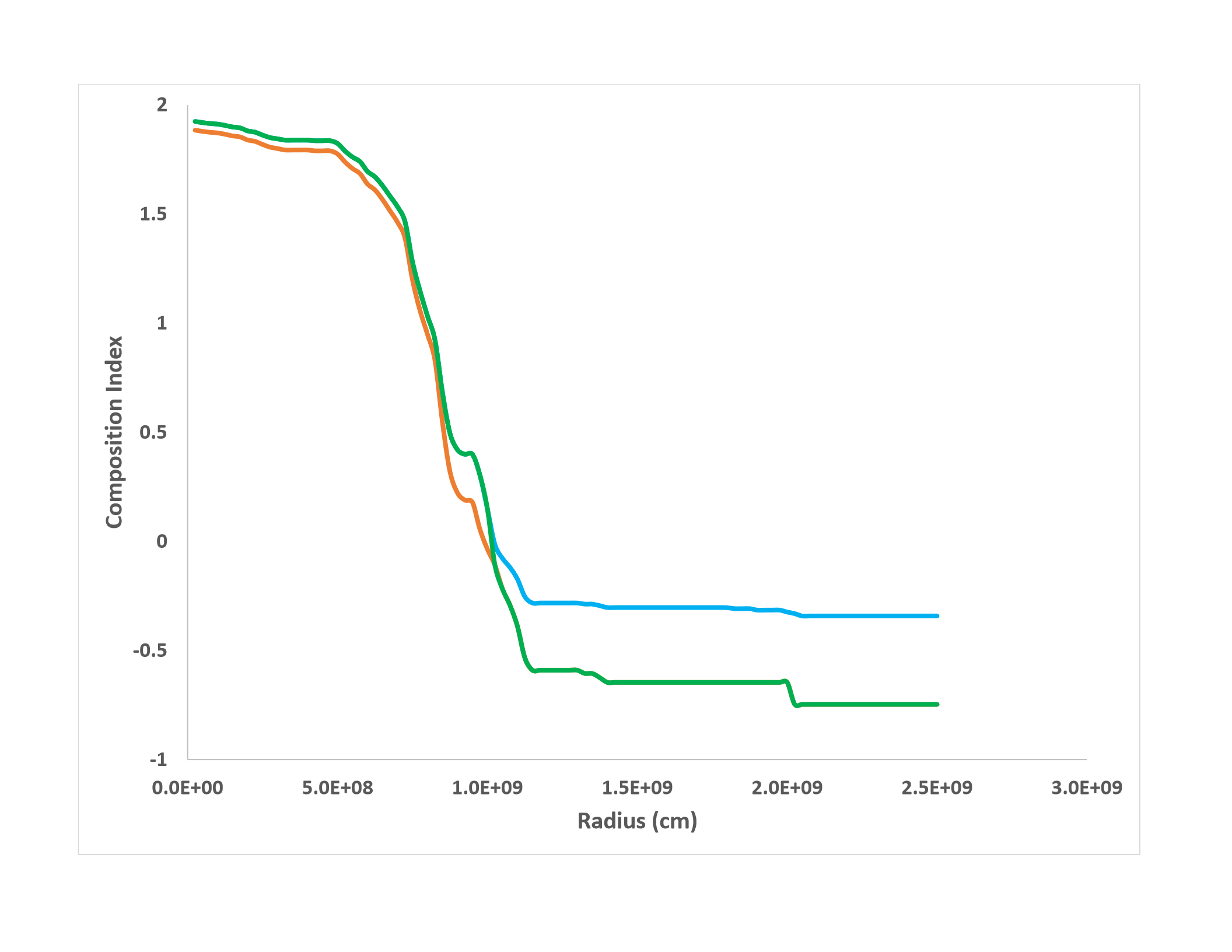}
	\caption{Composition profile, based on the bi-directional algorithm, for the density distribution in Fig.\,\ref{fig:random_model_example}. The composition is given by an index computed from the linear combination of the mass fraction times the index for each material, -1=pure envelope; 0=pure water; 1=pure rock and 2=pure iron. {\bf Left:} Composition index derived from the inward (orange) and outward (blue) algorithms. {\bf Right:} Same as the left panel, but with the index derived from the bi-directional algorithm (green) superimposed.}
	\label{fig:bidirectional_selection}
\end{figure}

\begin{figure}[h!]
	\centering
	\includegraphics[height=5cm]{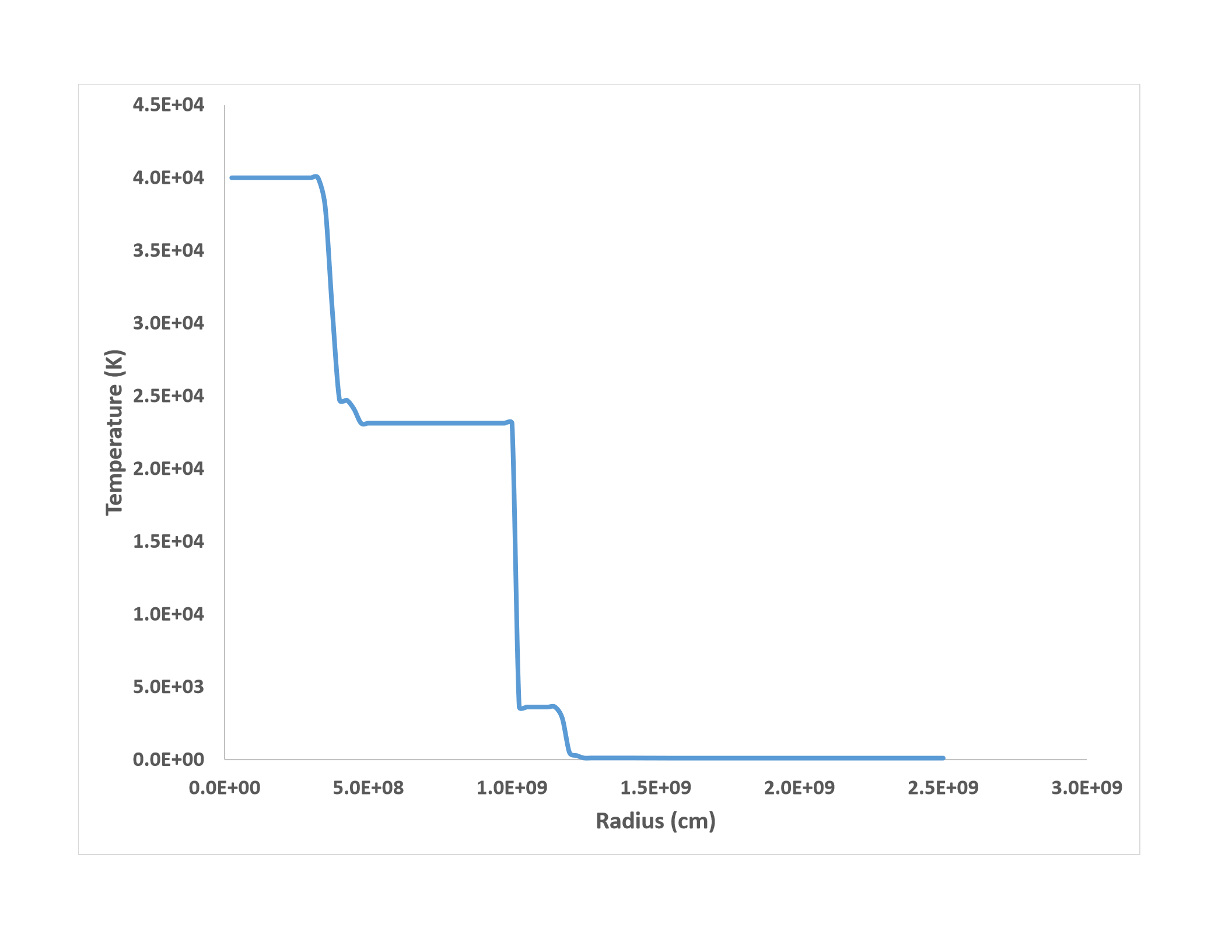}
	\hspace{1cm}
	\includegraphics[height=5cm]{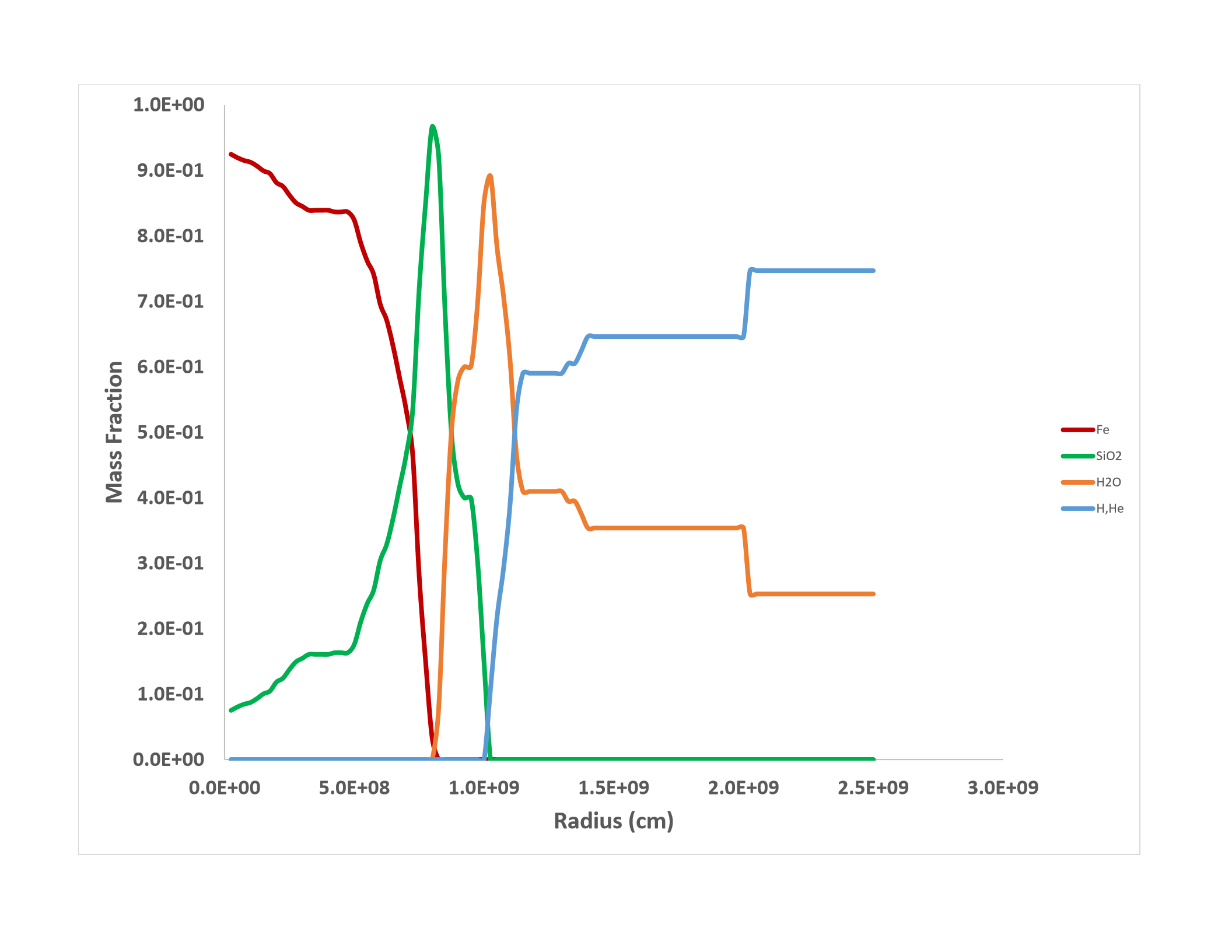}
	\caption{Temperature and composition profiles for the density distribution of Fig.\,\ref{fig:random_model_example} as derived from the bi-directional algorithm. {\bf Left:} Temperature as a function of radius. {\bf Right:} Mass fraction of Fe (red), \sio2 (green), water (orange), and envelope (blue) as a function of radius.}
	\label{fig:bidirectional}
\end{figure}

Finally, the bi-directional algorithm is a blend of the two previous methods. The idea is to both maximize the rock/water mass ratio, and also allow for compositions that have significant envelopes and are thus more appropriate for gaseous planets.  For each shell, the algorithm simply compares between the inward and outward compositions of the same random density model, and chooses whichever has less water in it. The result is shown in Fig.\,\ref{fig:bidirectional_selection} for the case of $T_c=4\times 10^4$\,K. Here we have plotted a single curve where the composition is denoted by an index given by a mass-weighted average of the indices of the different material pairs.  The indices are 2 for Fe, 1 for \sio2, 0 for water, and -1 for envelope. 
Thus a value between 1 and 2 indicates a mixture of Fe and \sio2, a value between 0 and 1, a mixture of \sio2 and water, and a value between 0 and -1, a mixture of water and envelope.  The left panel overplots the inward and outward algorithms.  The right panel shows how to generate the bi-directional curve by minimizing water.  The resulting temperature and composition profiles are shown in Fig. \ref{fig:bidirectional}.  This model contains 5.3 $M_{\oplus}$ of rock and 3.9 $M_{\oplus}$ of water for a rock to water ratio of 1.4.

\begin{figure}[h!]
	\centering
	\includegraphics[height=6cm]{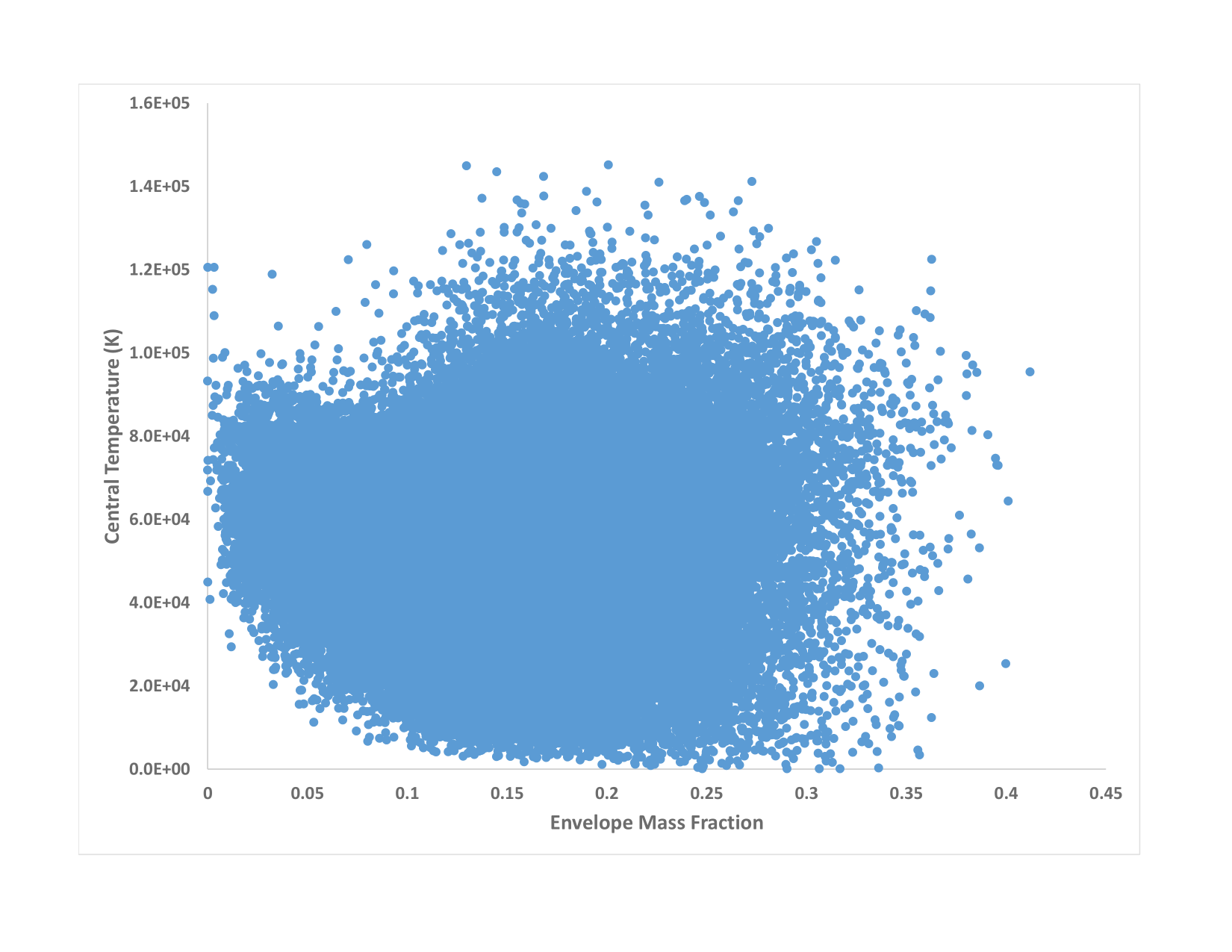}
	\includegraphics[height=6cm]{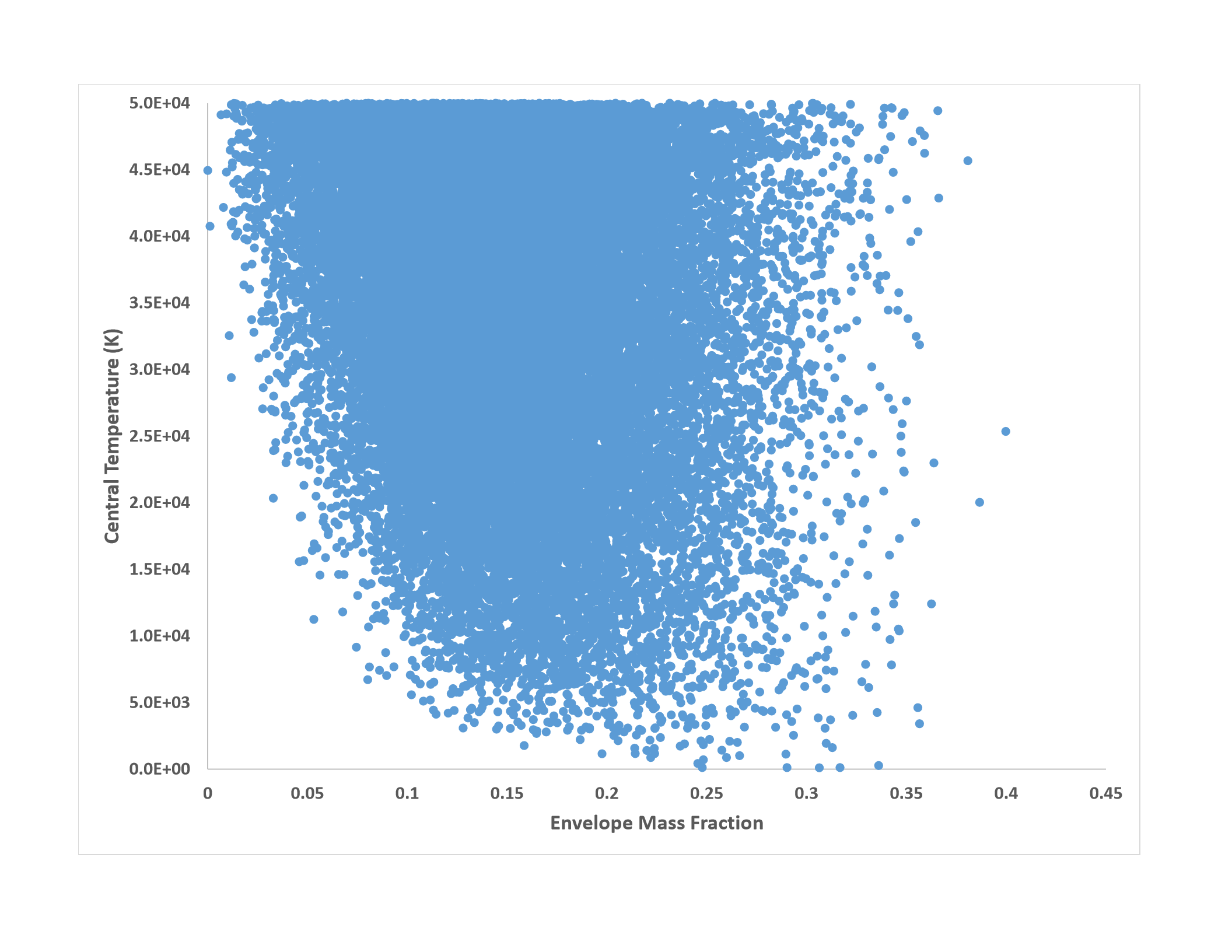}
	\includegraphics[height=6cm]{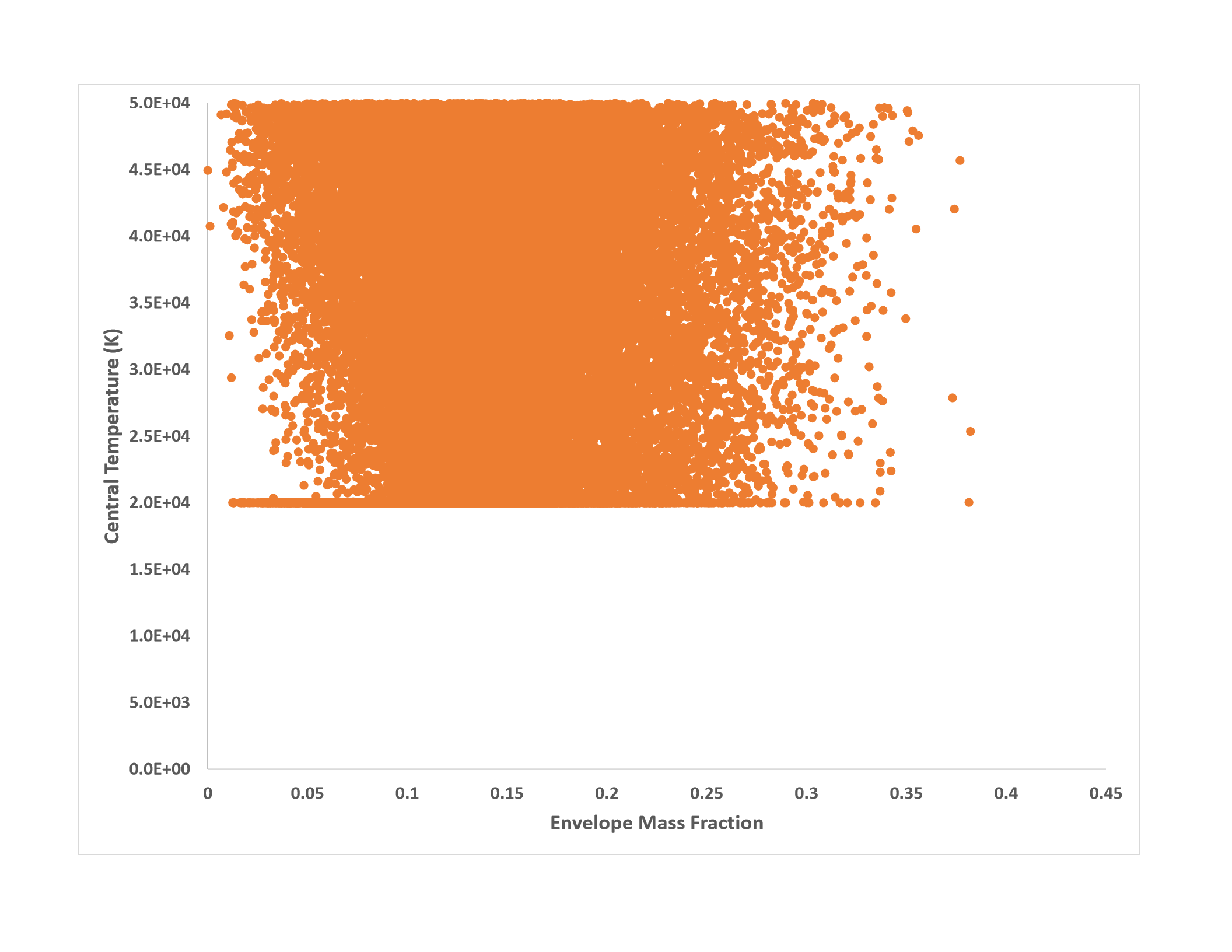}
	\includegraphics[height=6cm]{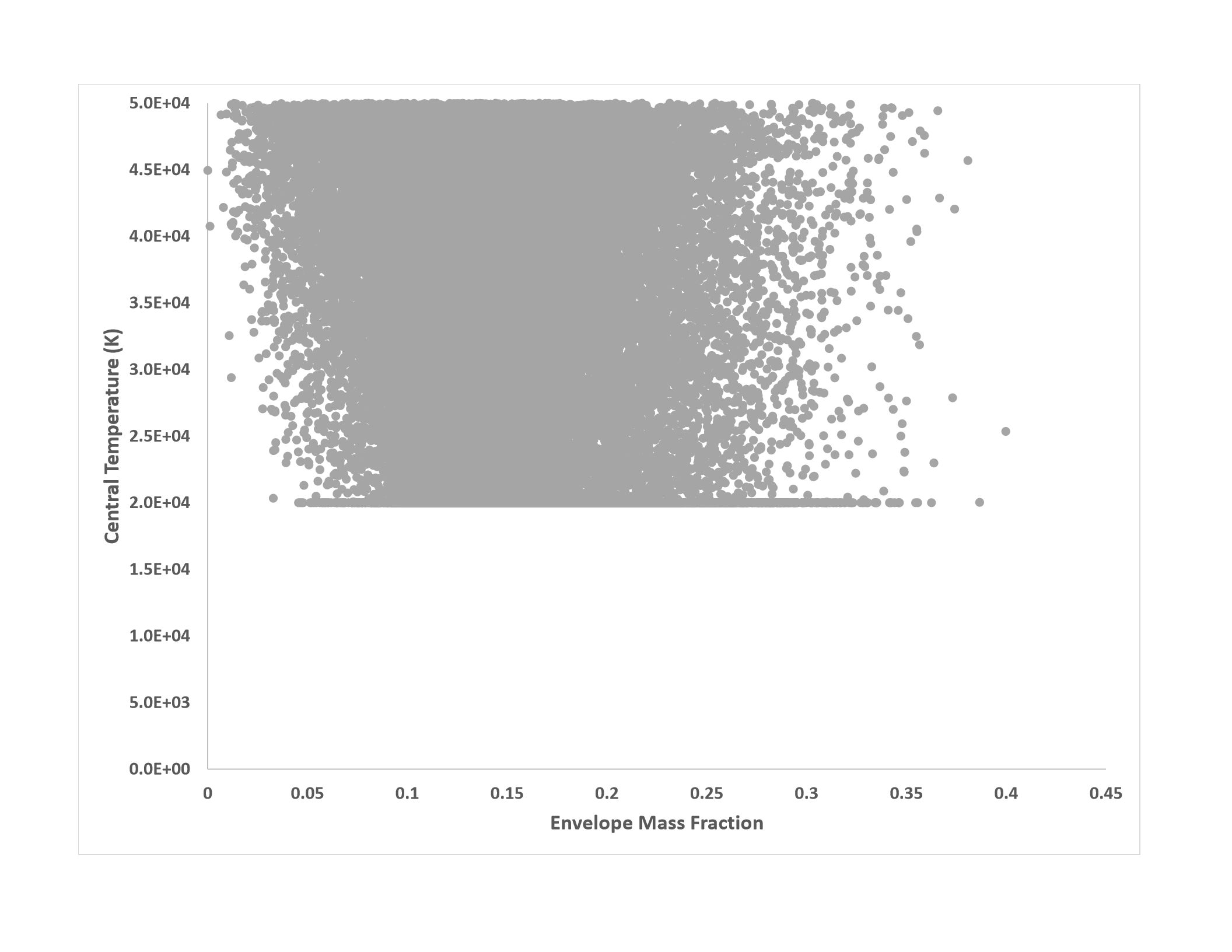}
	\caption{Envelope mass fraction vs. central temperature for 100,000 models generated with the different algorithms.  Each dot is a different model.  {\bf Upper Left:} Full set of models for the inward algorithm.  {\bf Upper Right:} Subset of models for the inward algorithm for $T_c\leq 5\times 10^4$\,K.  {\bf Lower Left:} Subset of models for the outward algorithm with $2\times 10^4$\,K$\leq T_c\leq 5\times 10^4$\,K.  {\bf Lower Right:} Subset of models for the bi-directional algorithm with $2\times 10^4$\,K$\leq T_c\leq 5\times 10^4$\,K.  See text for details.}
	\label{fig:et}
\end{figure}

\section{Results}\label{S:Results}
	\begin{figure}[h!]
	\centering
	\includegraphics[width=8cm]{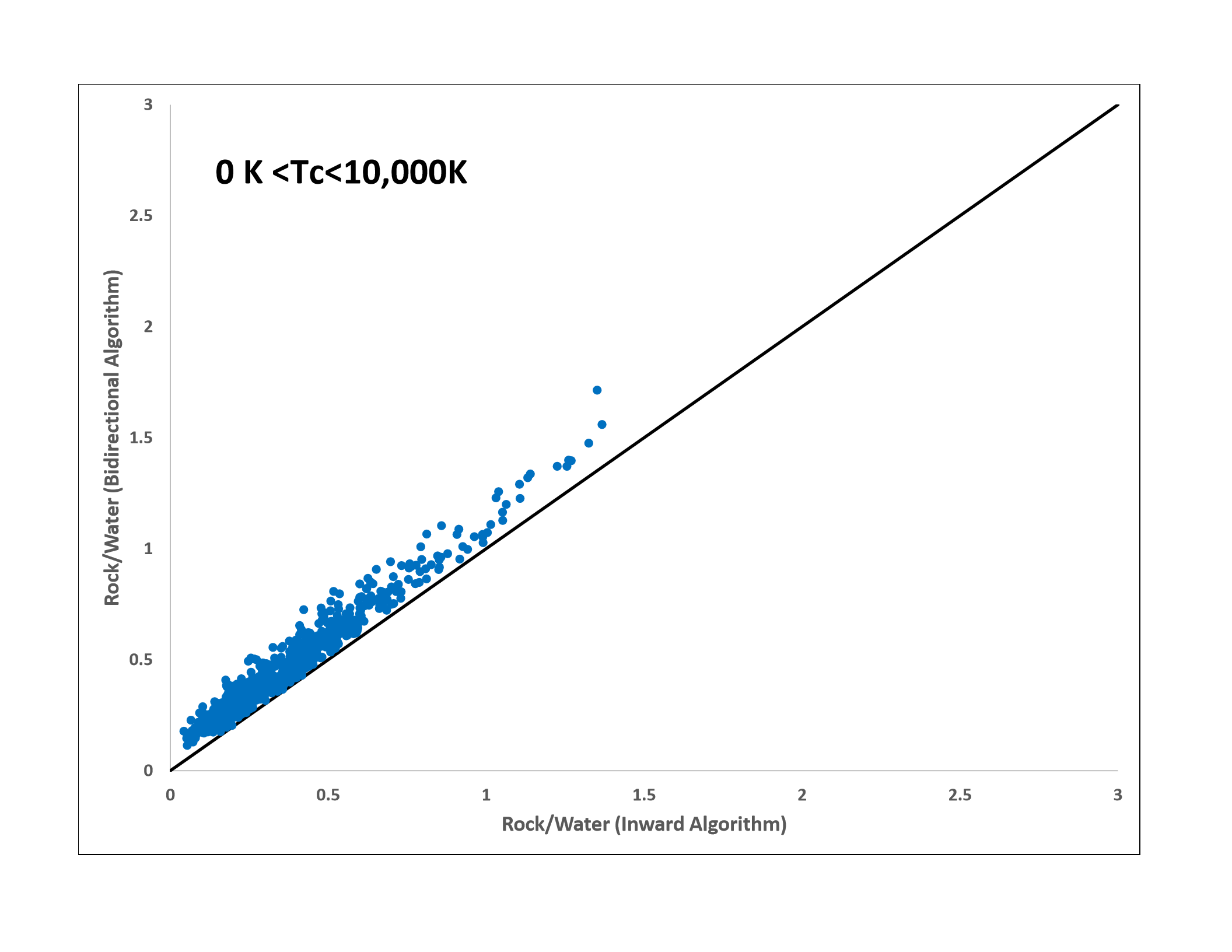}
	\includegraphics[width=8cm]{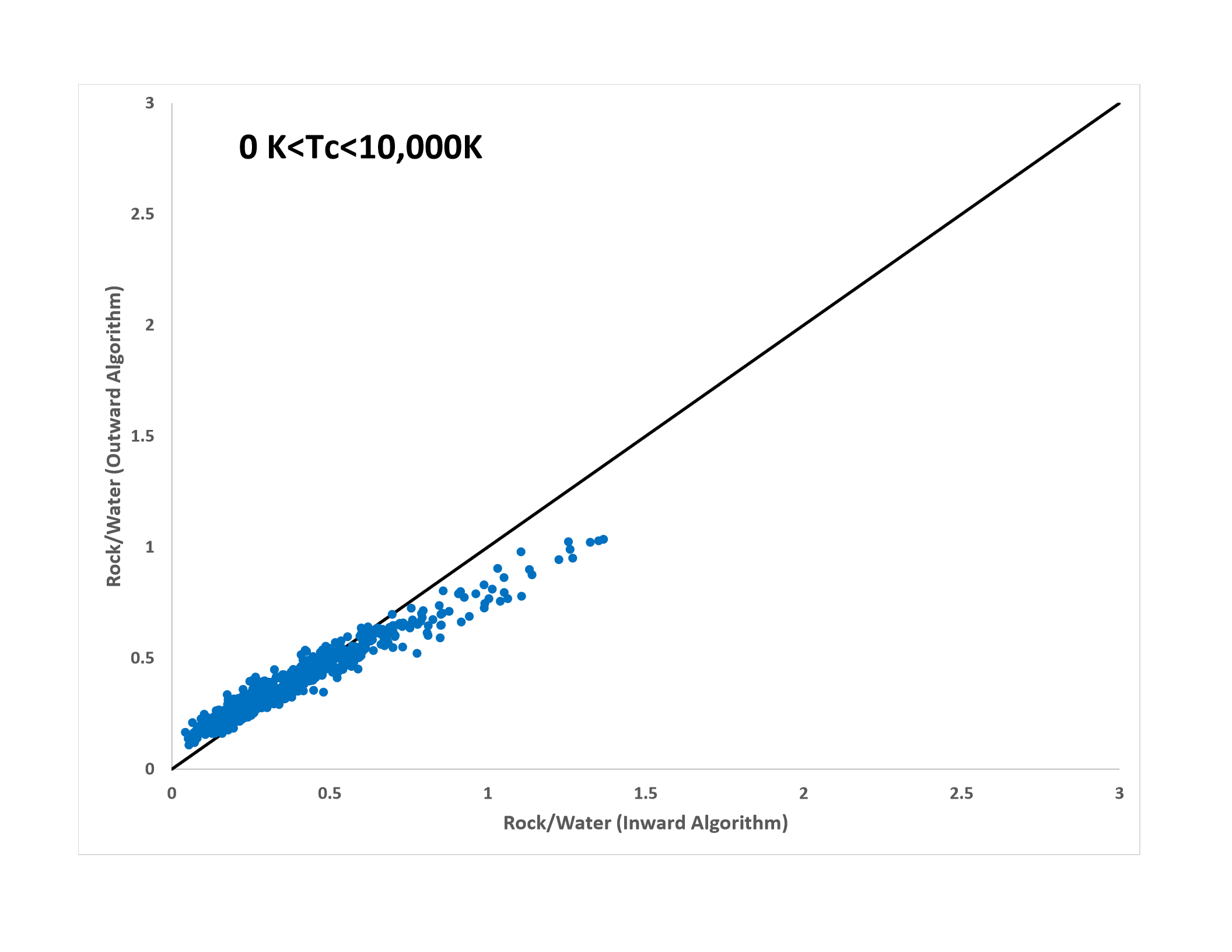}
	\includegraphics[width=8cm]{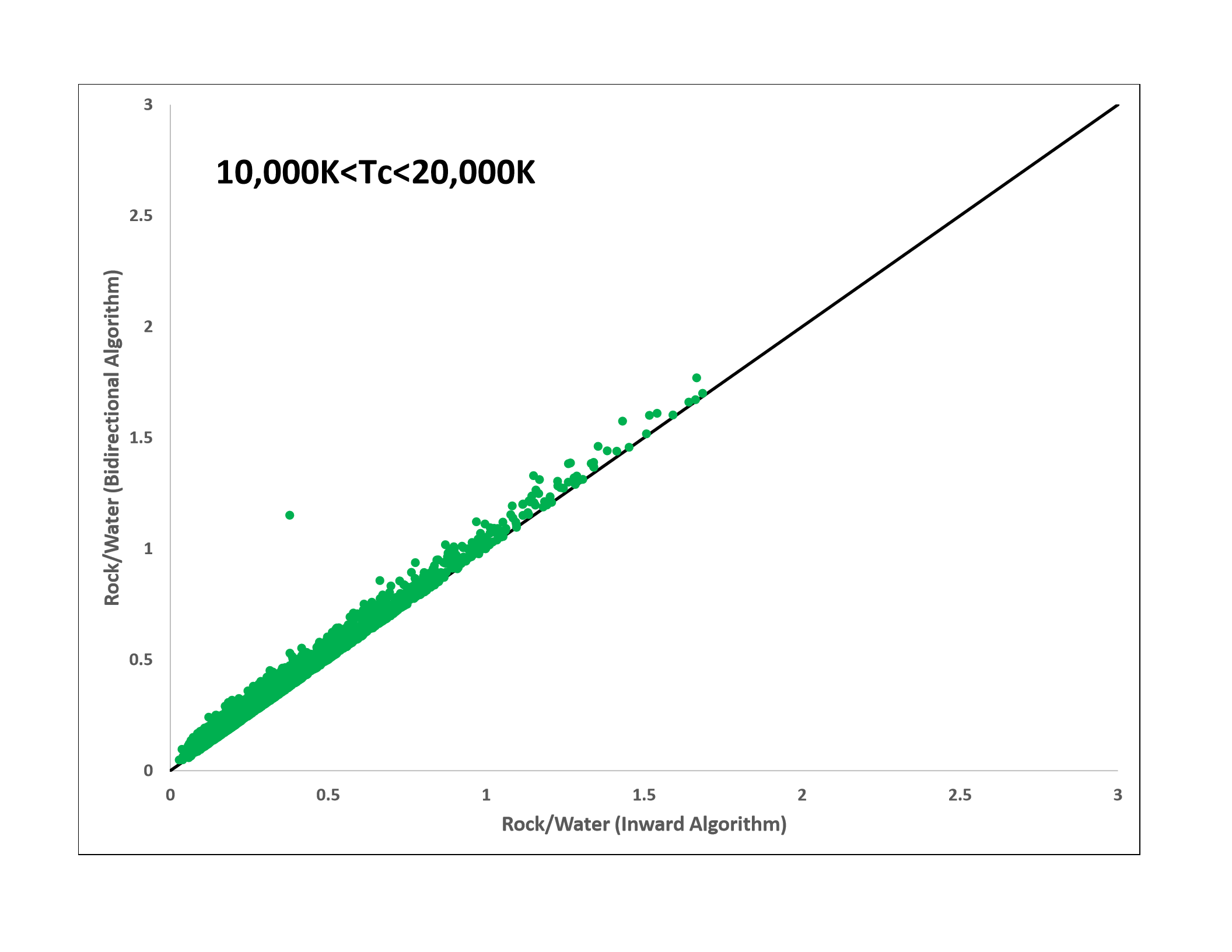}
	\includegraphics[width=8cm]{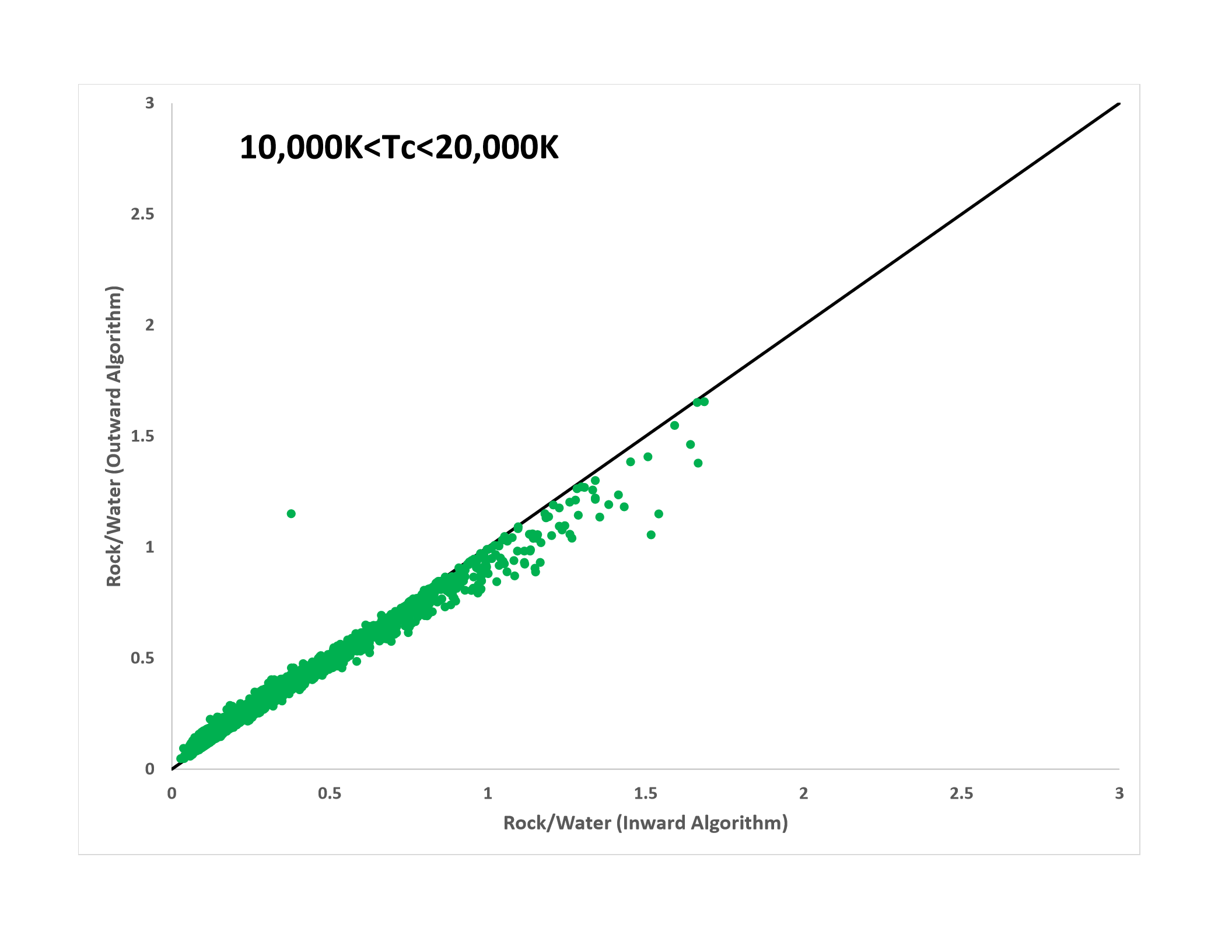}
	\includegraphics[width=8cm]{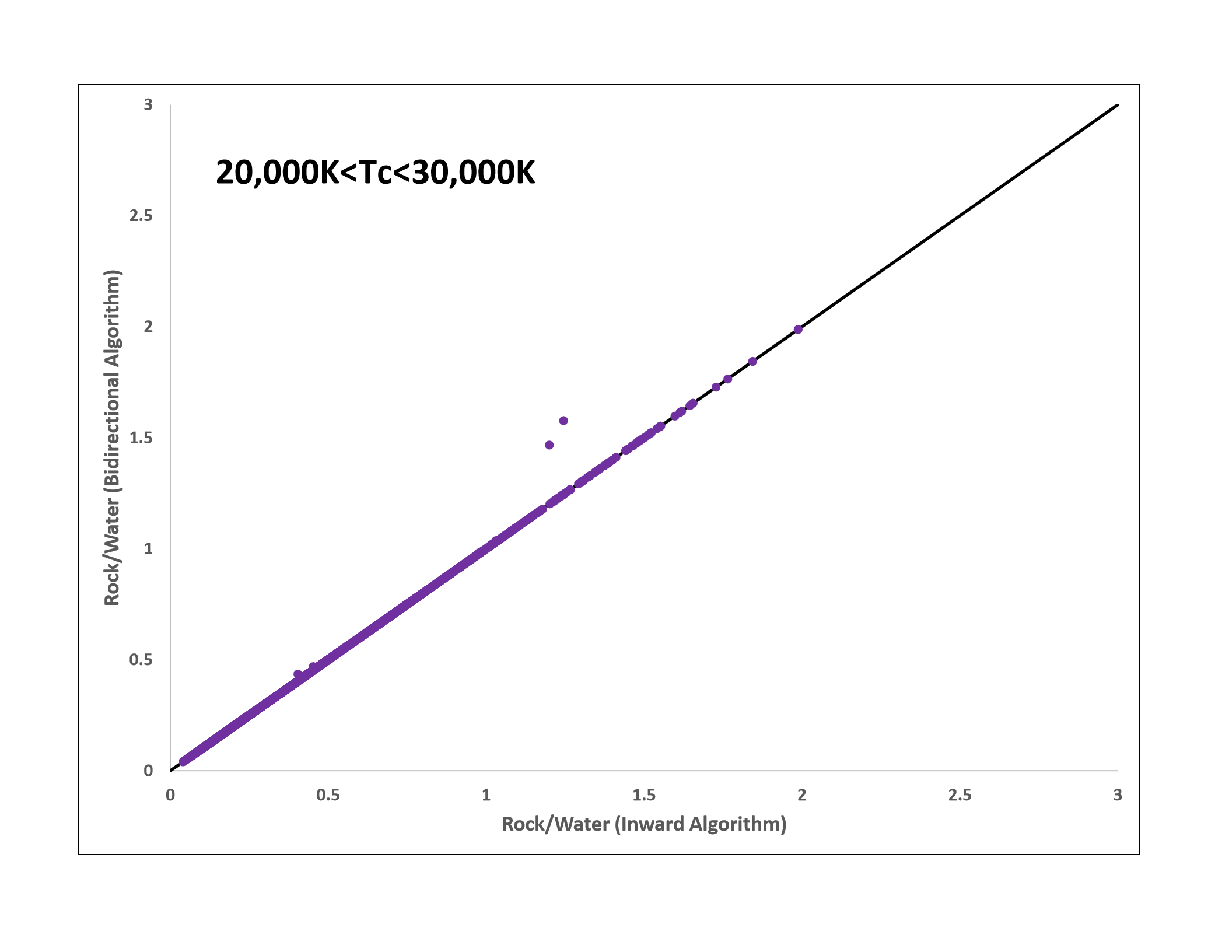}
	\includegraphics[width=8cm]{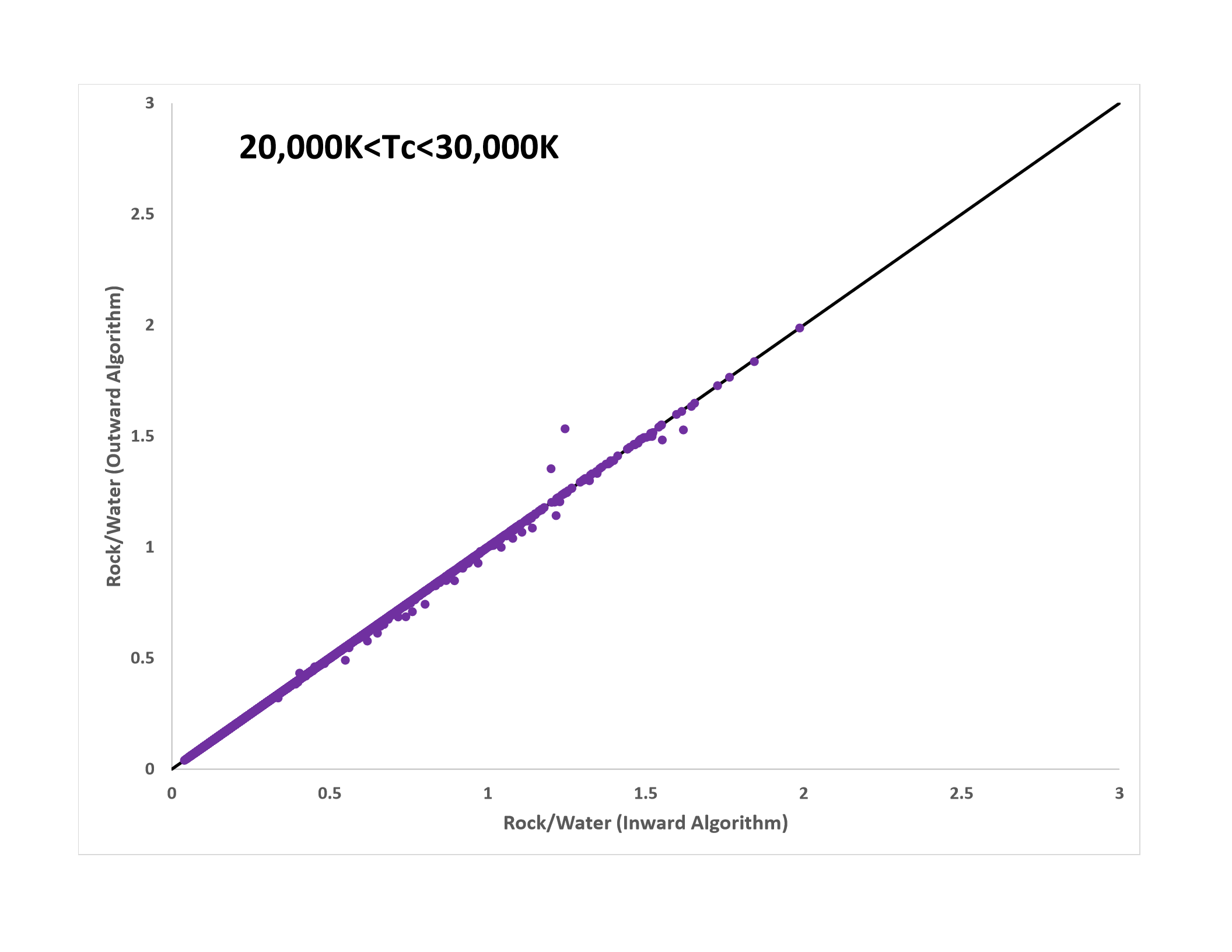}
	\caption{{\bf Left} Rock to water ratio derived from the bi-directional algorithm (y-axis) as a function of the rock to water ratio from the inward algorithm (x-axis) for 3 different ranges of central temperature. {\bf Right} Rock to water ratio derived from the outward algorithm (y-axis) as a function of the rock to water ratio derived from the inward algorithm (x-axis) for the same 3 temperature ranges.  The black line shows where y=x.}
	\label{fig:rwzoom}
\end{figure}

\begin{figure}[h!]
	\centering
	\includegraphics[width=8cm]{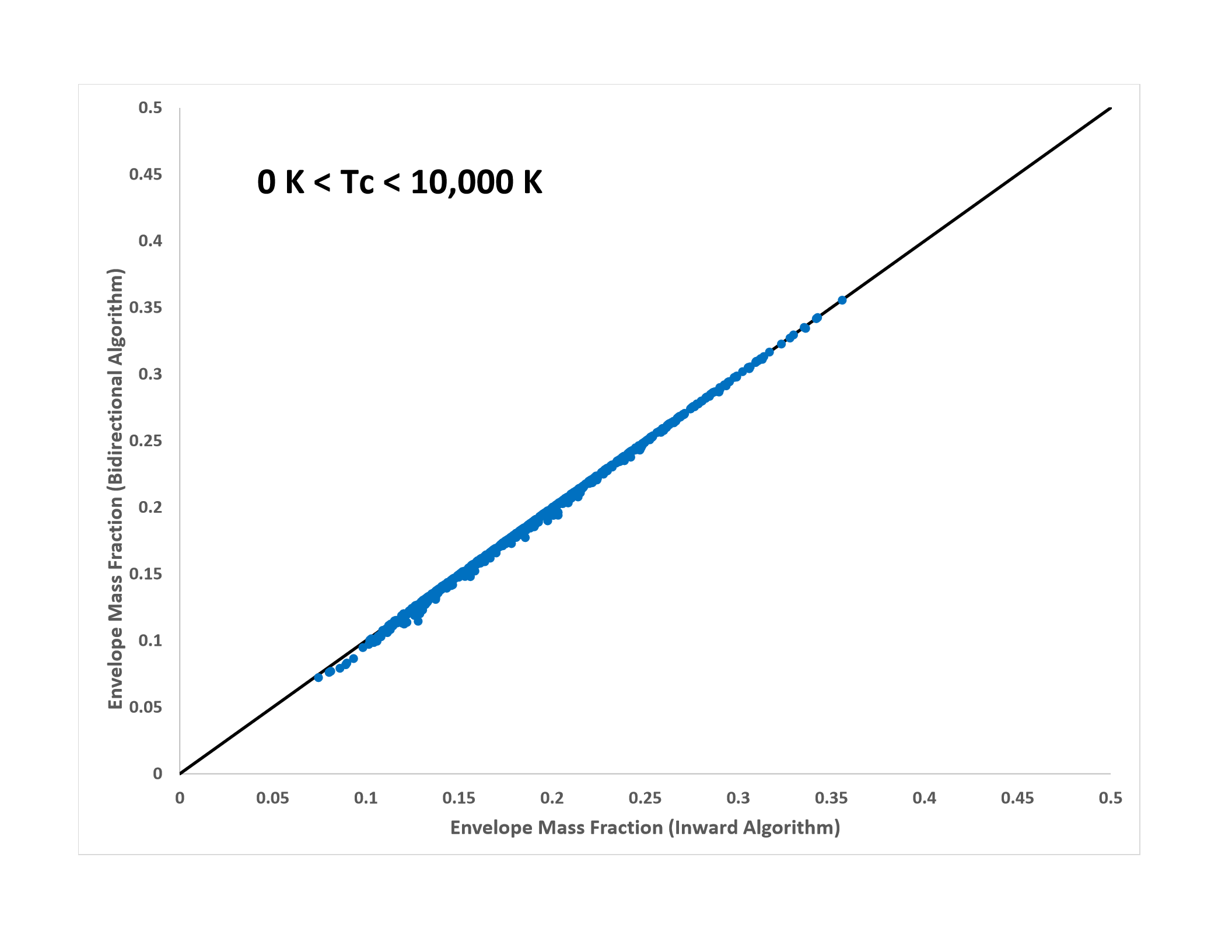}
	\includegraphics[width=8cm]{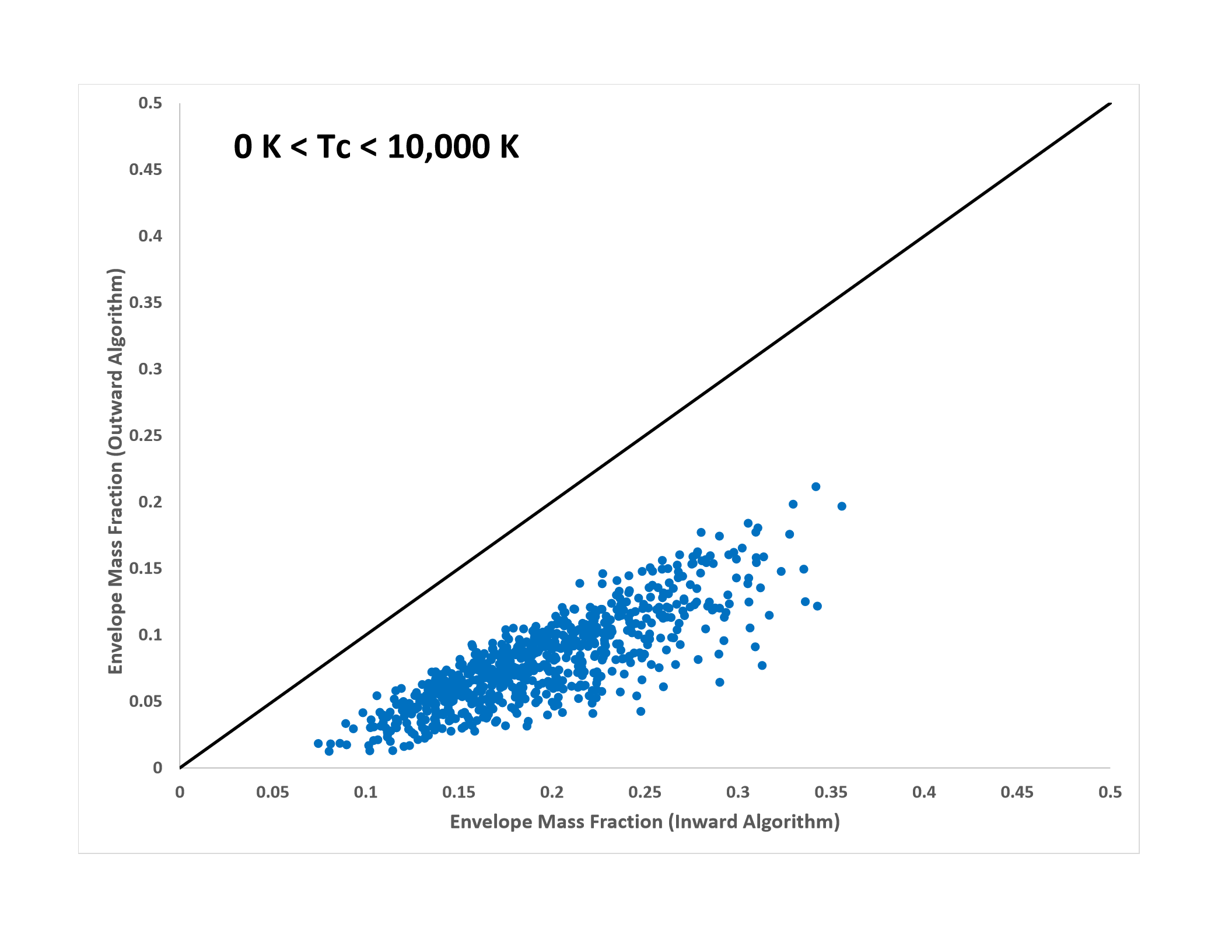}
	\includegraphics[width=8cm]{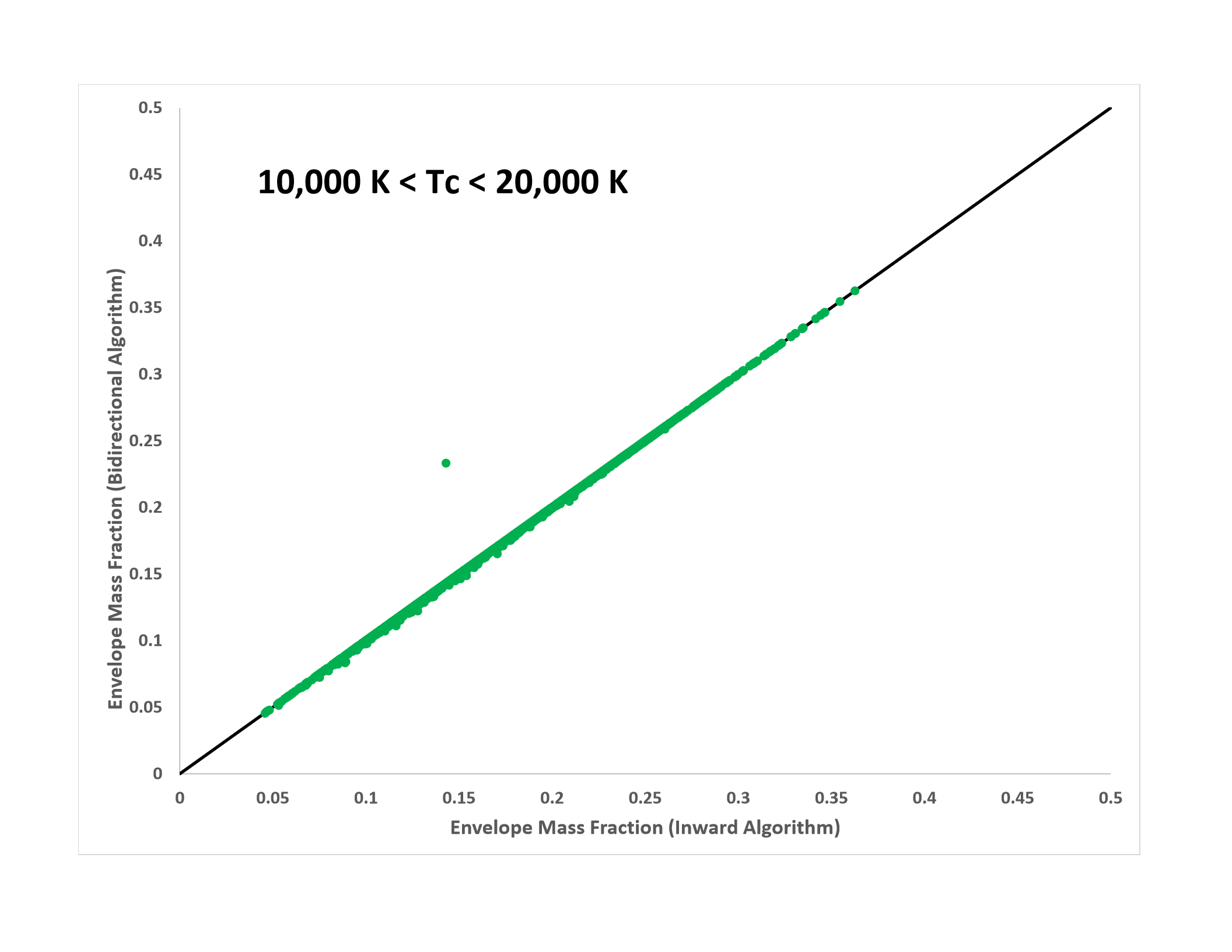}
	\includegraphics[width=8cm]{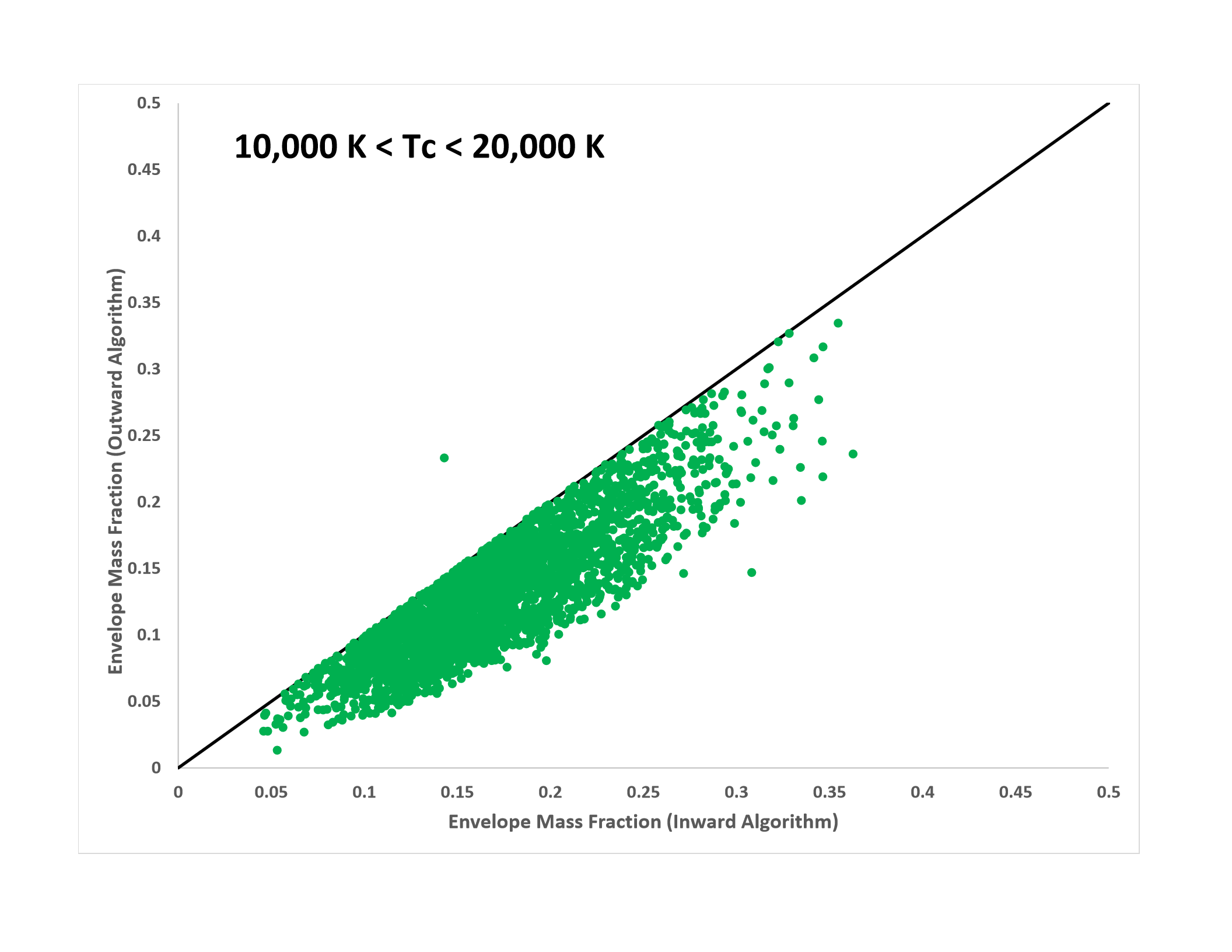}
	\includegraphics[width=8cm]{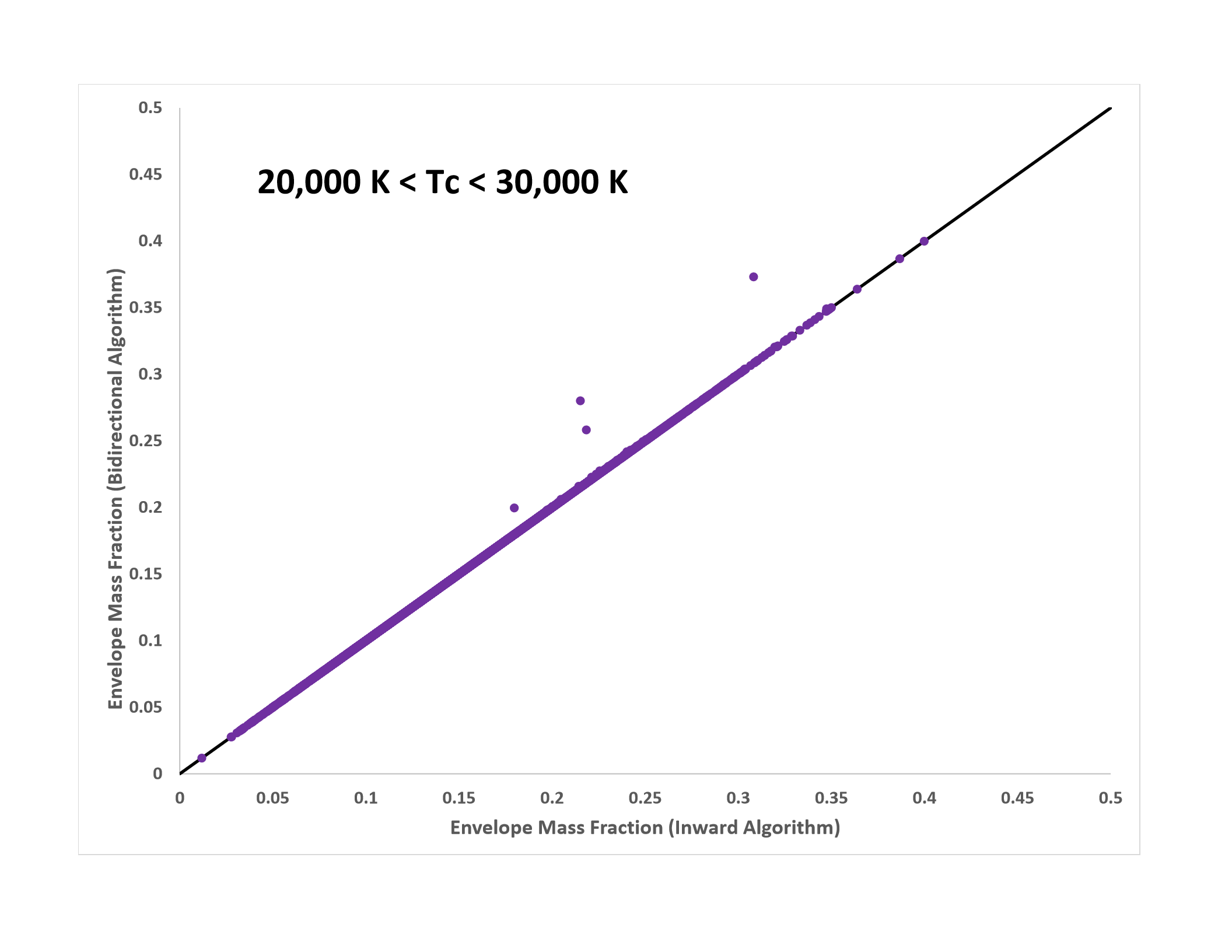}
	\includegraphics[width=8cm]{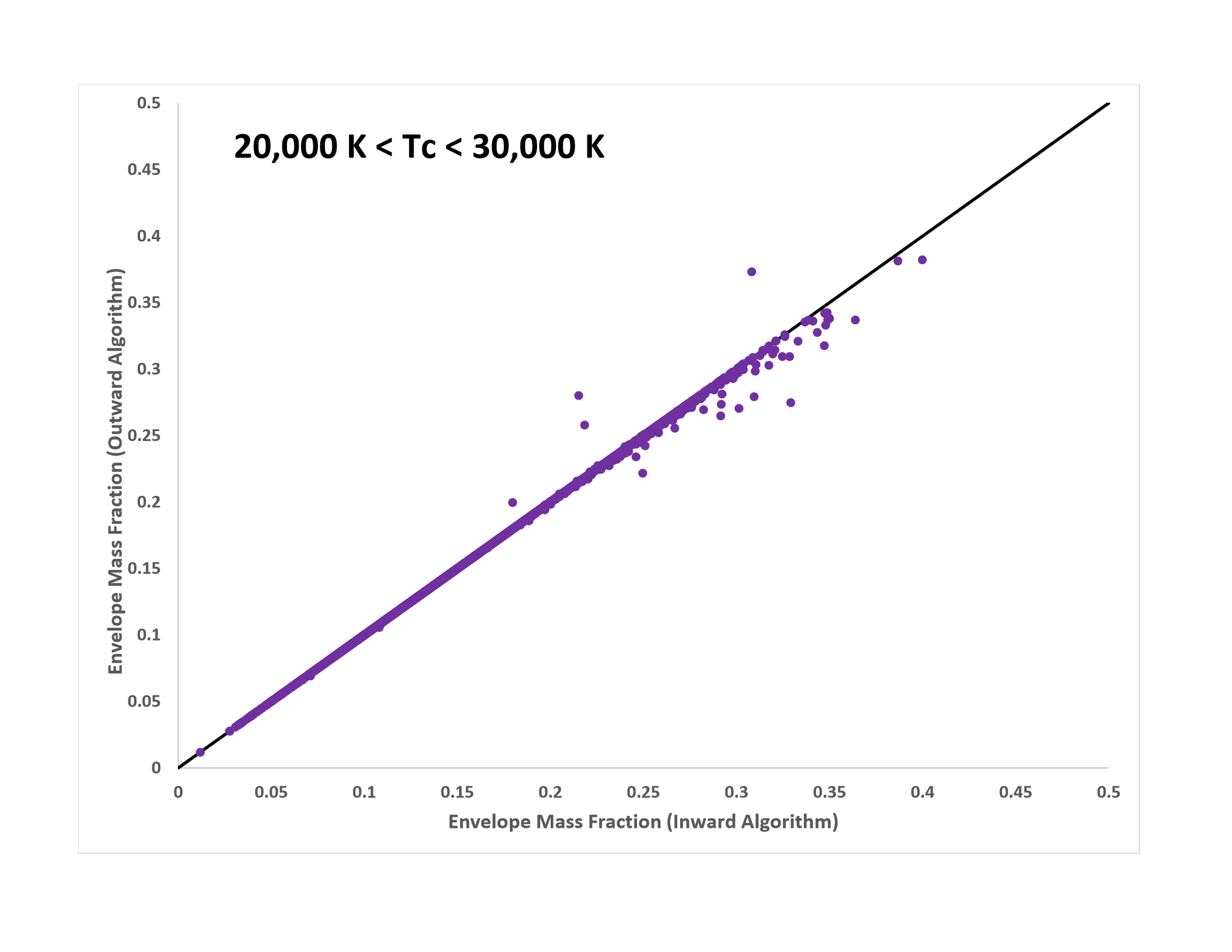}
	\caption{{\bf Left} Envelope mass fraction derived from the bi-directional algorithm (y-axis) as a function of the envelope mass fraction from the inward algorithm (x-axis) for 3 different ranges of central temperature. {\bf Right} Envelope mass fraction derived from the outward algorithm (y-axis) as a function of the envelope mass fraction derived from the inward algorithm (x-axis) for the same 3 temperature ranges.  The black line shows where y=x.}
	\label{fig:env}
\end{figure}

In all, we ran 100,000 models.  The inward algorithm produced 74,092 successful models, while the outward and bi-directional algorithms each produced 73,940 models.  For both the outward and bi-directional algorithms we assumed $T_c\geq 2\times 10^4$\,K.  

The upper left panel of Fig.\,\ref{fig:et} shows the envelope mass fraction and central temperature for the models computed using the inward algorithm.  As can be seen, some of the models have central temperatures that are over $10^5$\,K, and we must ask what a reasonable upper limit is for Uranus.  Classical models, based on the assumption of an adiabatic temperature profile typically yield $T_c\lesssim 10^4$\,K \citep{podcam74,helledetal10,redmer11}.  However, more recent works consider the possibility of processes that inhibit convection and yield central temperatures of $2\times 10^4$\,K and above \citep{podolak2019,vazanhel2020,helled2020}.  For the purposes of our study, we take an upper limit of $5\times 10^4$\,K.  Uranus' true central temperature is almost certainly less than this, but we set the limit purposely high in order to see the effect of such high temperatures. 33,440 of the models computed with the inward algorithm, and 33,362 each with the outward and bi-directional algorithms fall within this range.

One feature that stands out is that the minimum envelope mass fraction decreases with increasing central temperature.  This is expected, since as the temperature increases the densities of the individual materials tend to decrease, and less envelope material is required to match Uranus' overall density.  On the other hand, the maximum envelope mass fraction rarely exceeds 0.4, independent of central temperature, and this seems to be an upper limit.   

The upper right panel of the figure shows a close-up of the region of interest.  This can be compared to the panel in the lower left, which shows the same models with the composition derived using the outward algorithm, taking the minimum central temperature to be $T_c=2\times 10^4$\,K.  The behavior is similar for the two cases except for central temperatures of $\sim 2\times 10^4$\,K and below.  Those models found with the inward algorithm with $T_c<10^4$\,K, will be hotter throughout when computed with the outward algorithm, and will therefore require a lower envelope mass fraction to match Uranus' overall density.  This accounts for the crowding of models near $T_c=2\times 10^4$\,K and the large spread of values for the envelope mass fraction.  

The lower right panel shows the same case with compositions computed using the bi-directional algorithm.  Here too there is a crowding around $T_c=2\times 10^4$\,K, but because of the bi-directionality of the algorithm, the crowding is less intense and shifted towards higher envelope mass ratios.

The consequence of applying the different algorithms can be gauged from Fig.\,\ref{fig:rwzoom}.  The left panels show the rock to water ratio derived from the bi-directional algorithm as a function of the same ratio as derived from the inward algorithm with that ratio.  The right panels show the same thing for the outward algorithm as a function of the inward algorithm.  The figure shows models for three inward algorithm temperature ranges, 0\,K$\leq T_c \leq 10^4$\,K (blue), $10^4$\,K$\leq 2\times 10^4$\,K (green), and $2\times 10^4$\,K$\leq 3\times 10^4$\,K (purple).  In most cases the points fall near a line with a slope of one, indicating that all three algorithms give similar results.  However, for the lowest $T_c$ range there is considerable scatter about that line, since both the bi-directional and the outward algorithms assume $T_c\geq 2\times 10^4$\,K, while the inward algorithm does not limit $T_c$.  As a result, the same density distribution in this range will always have a lower $T_c$ for the inward case than for the other two, and the compositions will differ somewhat.  The other thing to note is that, for the lowest $T_c$ range, although in the bi-directional case the points fall around a line with a slope of one, the points in the outward case fall around a line with a lower slope.  This is because the outward algorithm, in this $T_c$ range, tends to produce compositions that are more water-rich and with smaller envelopes.  These trends lessen for inward models with a higher $T_c$, and to practically disappear for $T_c\geq 2\times 10^4$\,K.

In Fig.\,\ref{fig:env} the left hand side shows the envelope mass fraction from the bi-directional algorithm as a function of the envelope mass fraction from the inward algorithm.  The right-hand side shows the same relation for the outward algorithm as a function of the inward algorithm.  As can be seen from the figure, the bi-directional and inward algorithms give similar values for the envelope mass fraction for all values of the inward $T_c$.  However the outward algorithm tends to give noticeably lower envelope mass fractions for inward values of $T_c\leq 2\times 10^4$\.K.  As before the different algorithms tend to converge for inward $T_c$ values above $2\times 10^4$\,K.

\begin{figure}[h!]
	\centering
	\includegraphics[height=6cm]{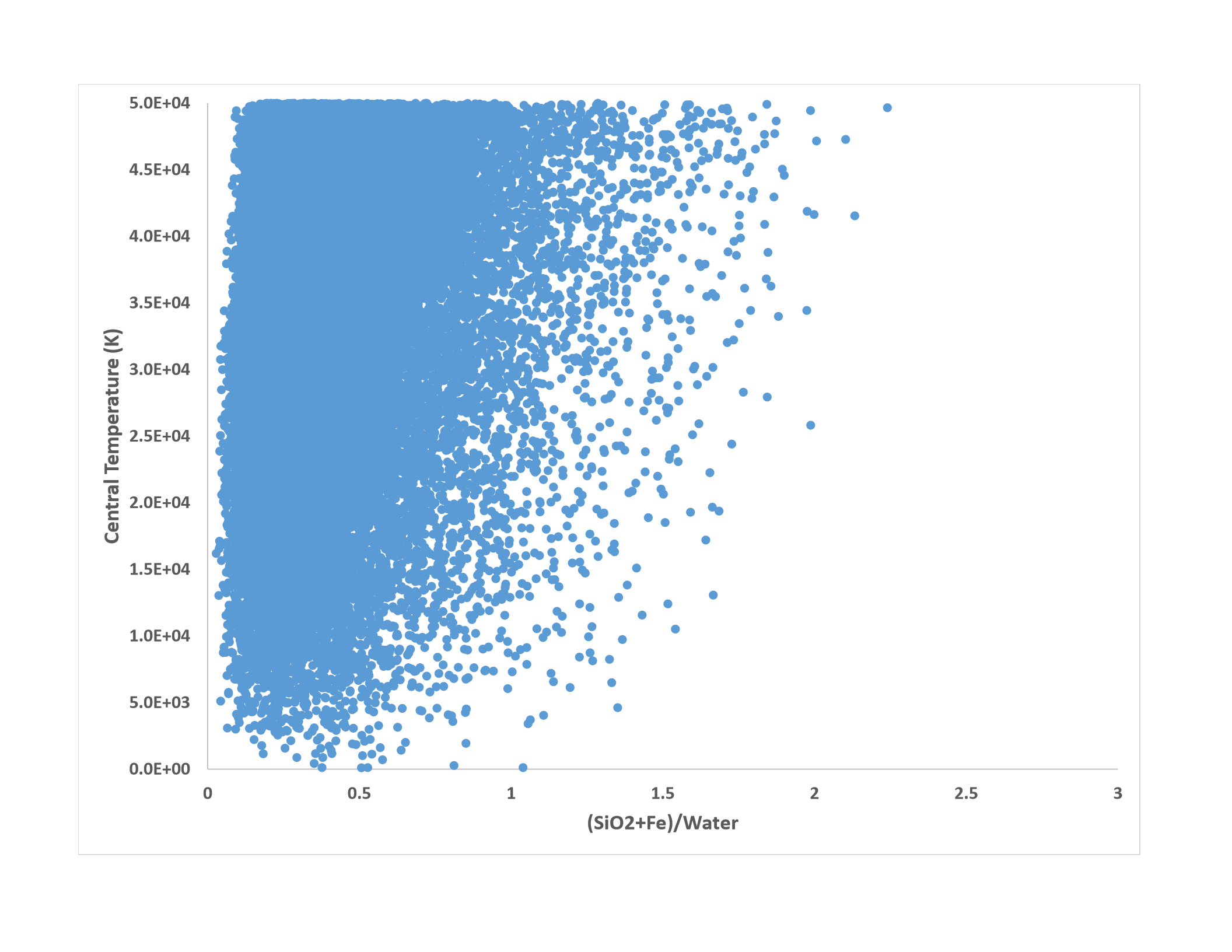}
	\includegraphics[height=6cm]{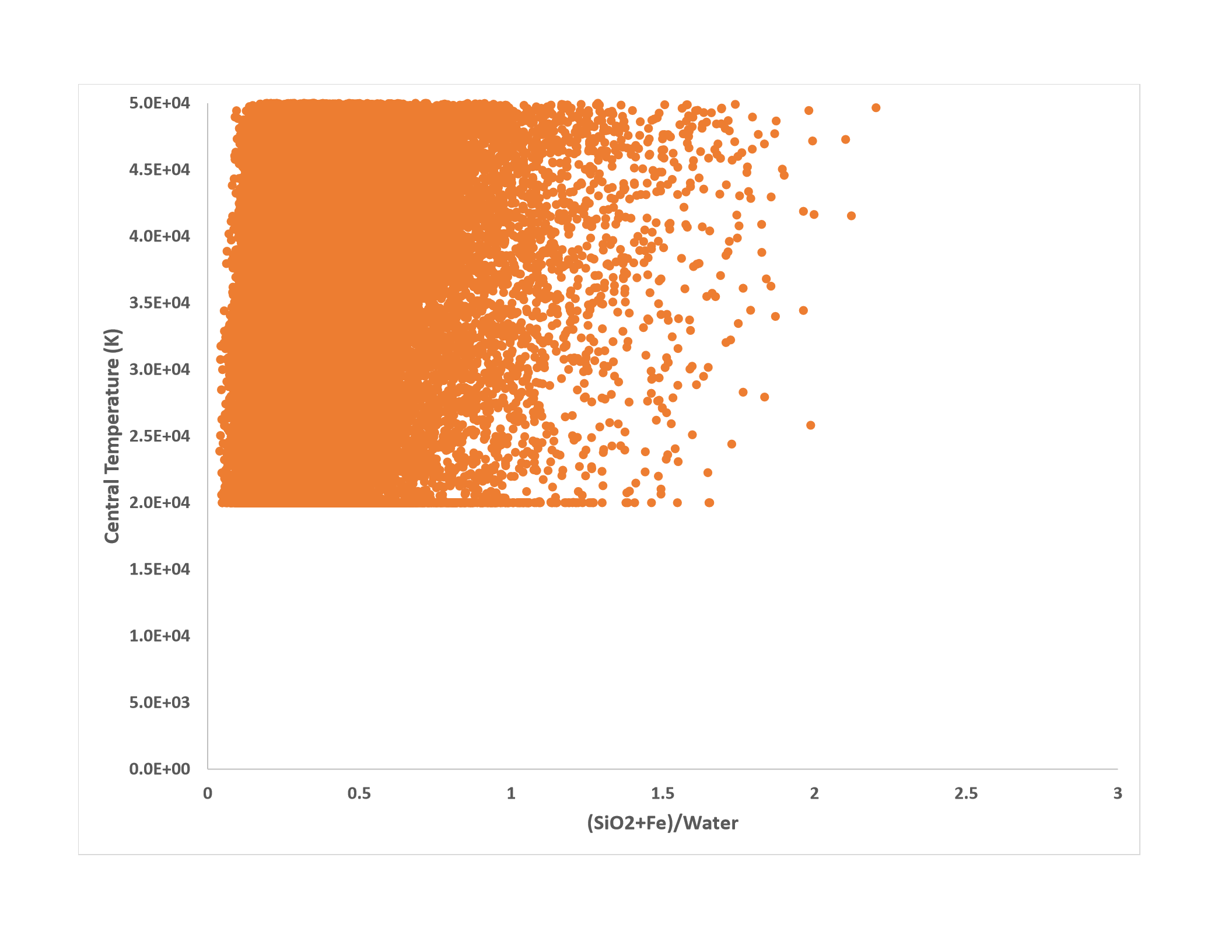}
	\includegraphics[height=6cm]{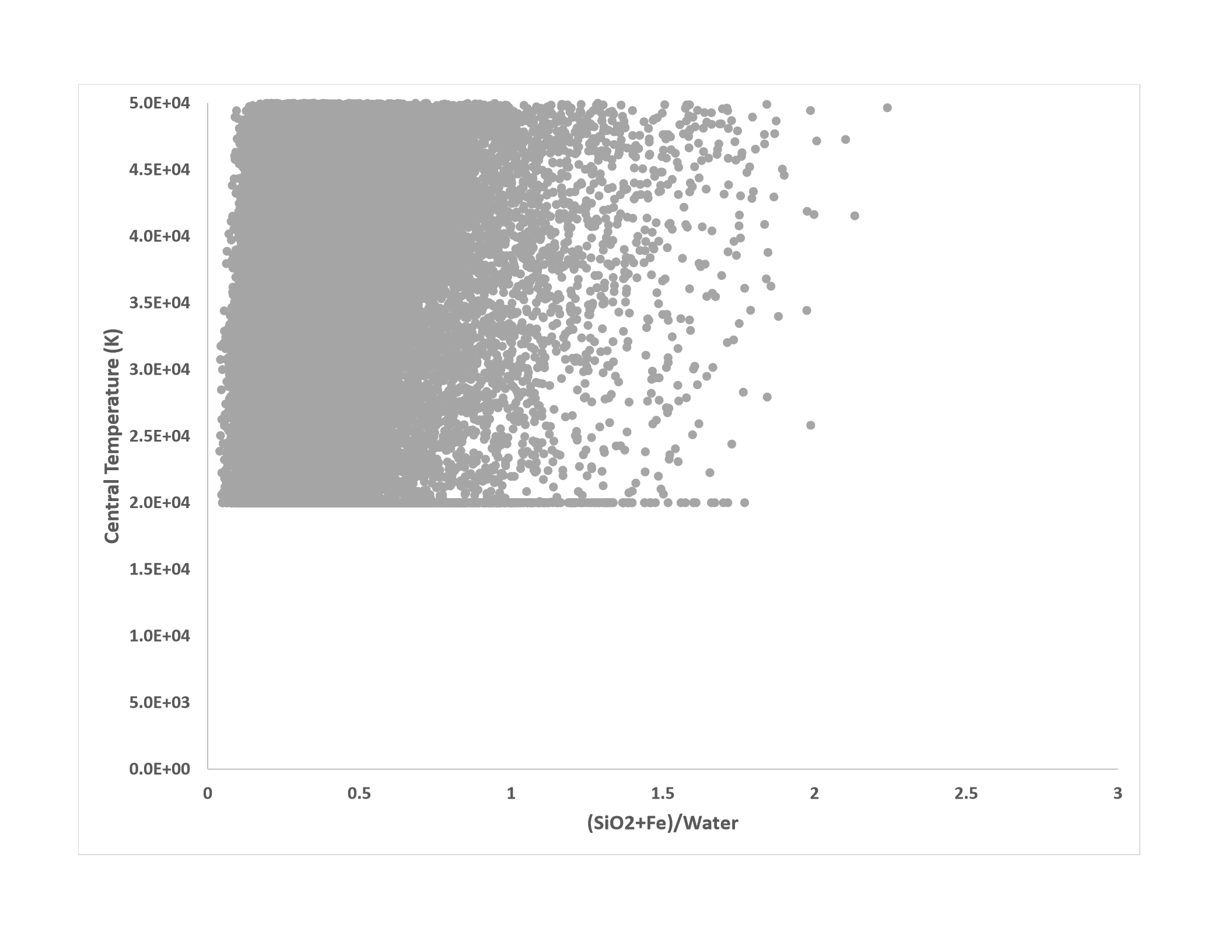}
	\includegraphics[height=6cm]{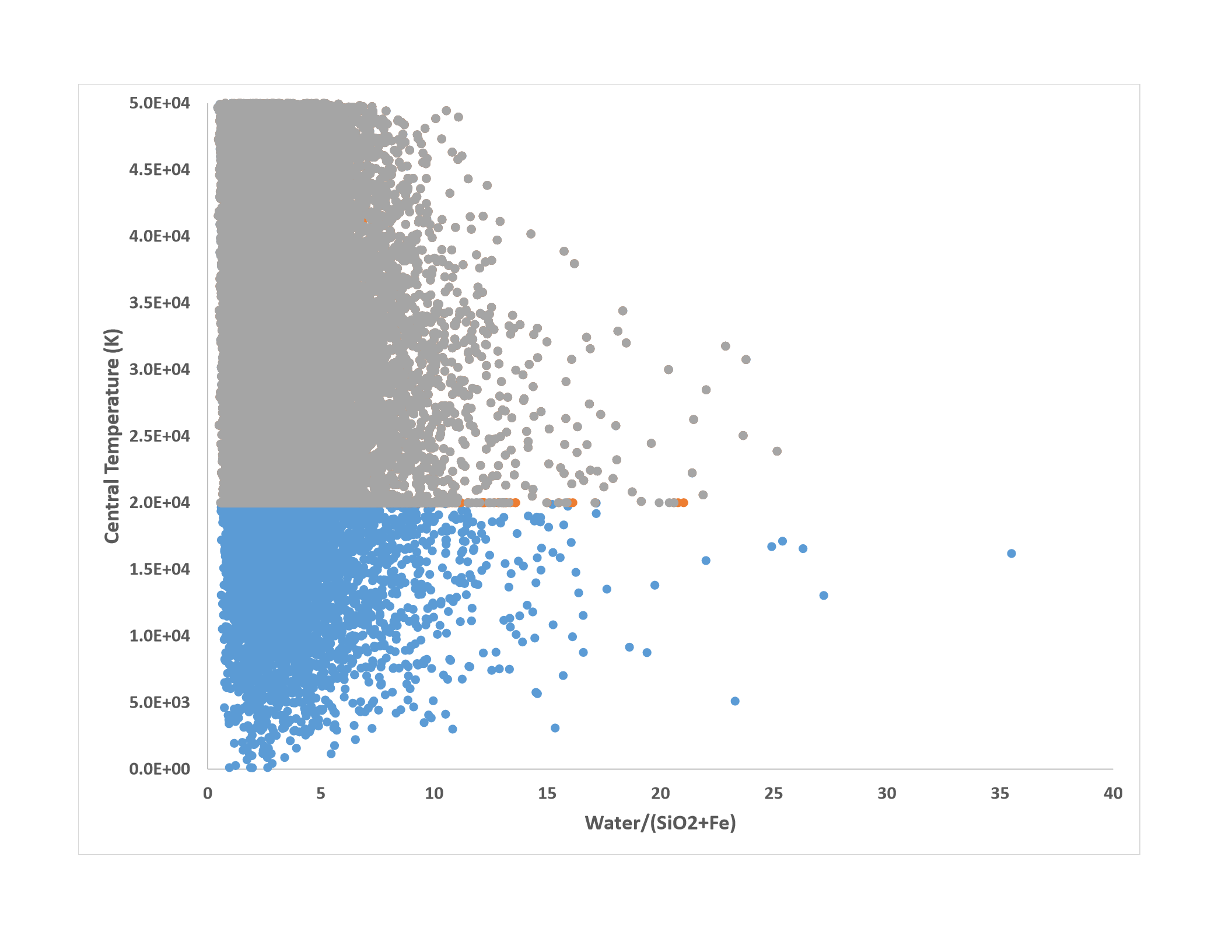}
	\caption{Rock to water mass ratio (x-axis) vs. central temperature (y-axis) for 100,000 models generated with the different algorithms discussed in the text.   {\bf Upper Left:} Inward algorithm for $T_c\leq 5\times 10^4$\,K.  {\bf Upper Right:} Outward algorithm for $T_c\leq 5\times 10^4$\,K.  {\bf Lower Left:} Bi-directional algorithm with $2\times 10^4$\,K$\leq T_c\leq 5\times 10^4$\,K.  {\bf Lower Right:} Overlay of the water to rock ratio (x-axis) vs. the central temperature for the inward algorithm (blue), outward algorithm (orange) and bi-directional algorithm (grey).}
	\label{fig:ri}
\end{figure}

Fig.\,\ref{fig:ri} shows the rock to water ratios derived from the different algorithms.  The upper left panel shows the case for the inward algorithm.  As the central temperature increases, larger rock to water ratios become possible, and we see that although they are rare, it is possible to find a model with a rock to water ratio of 2 if we allow the central temperature to reach $2.5\times 10^4$\,K.  A similar result was found by \cite{vazanhel2020}, but they required hydrogen and helium to be mixed all the way to the center of the planet.  In this model, the composition at the center is a mix of Fe and \sio2.  The models also show a slight increase in the minimum rock to water ratio with increasing $T_c$, as expected.  The upper right and lower left panels show the results for the outward and bi-directional algorithms, respectively.  The rock to water ratios for the different algorithms display very similar ranges and behaviors for all three algorithms.

In the lower right panel we have plotted, for all three algorithms, the inverse of the rock to water ratio (i.e. the water to rock ratio) in order to expand the region of low rock content.  Some models give quite high values, reminiscent of the models of \cite{podolak95}, but most, especially at the lower central temperatures, cluster around a water to rock ratio of $\sim 1-10$.  The expected solar water to rock ratio is roughly 2 \citep{Lodders2010}.

\begin{figure}[h!]
	\centering
	\includegraphics[height=10cm]{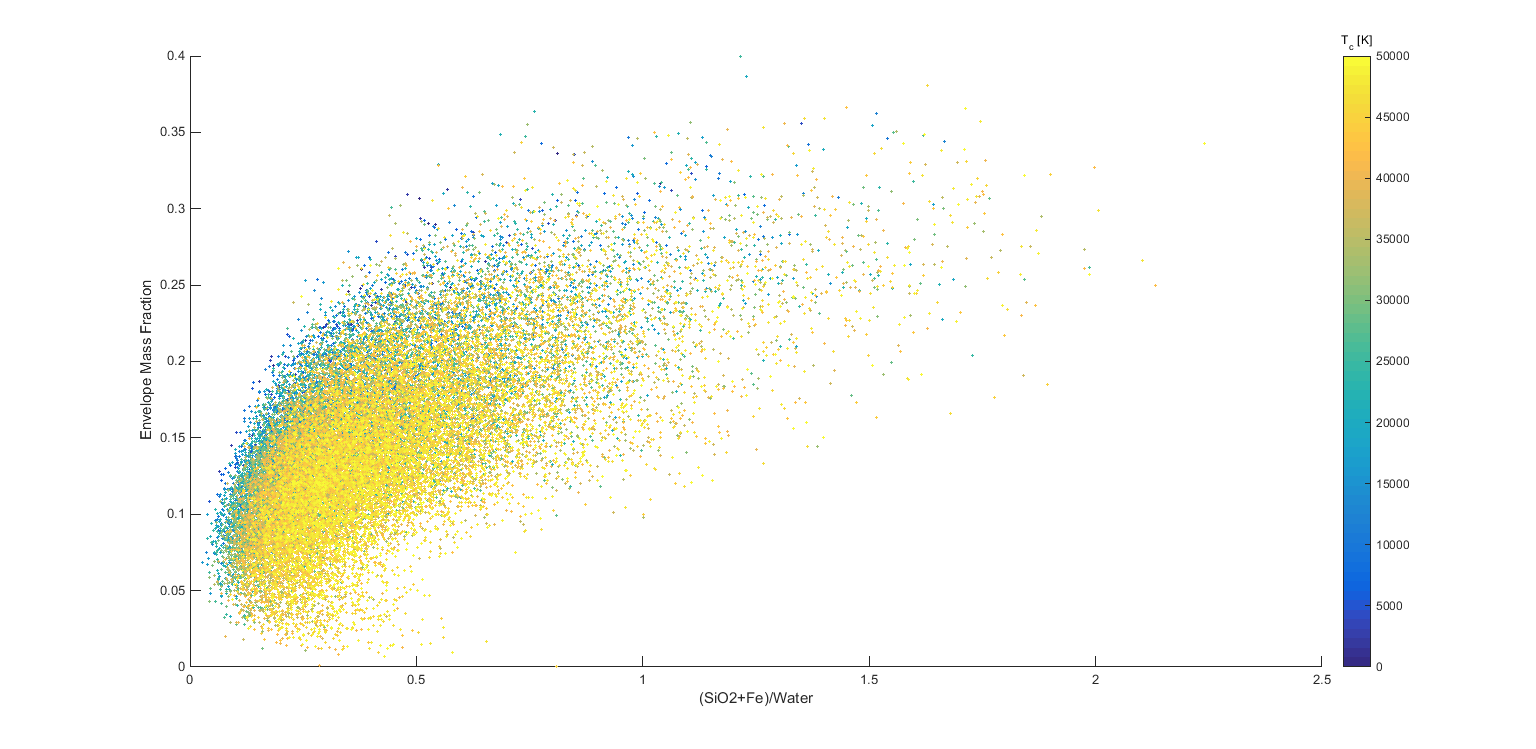}
	\caption{Envelope mass/total mass (y-axis) vs. rock to water ratio (x-axis) for models computed using the inward algorithm.  The color of the dots represents $T_c$ via the color bar on the right.  Shown are models with $T_c\leq 5\times 10^4$\,K.}
	\label{fig:er}
\end{figure}

\begin{figure}[h!]
	\centering
	\includegraphics[height=10cm]{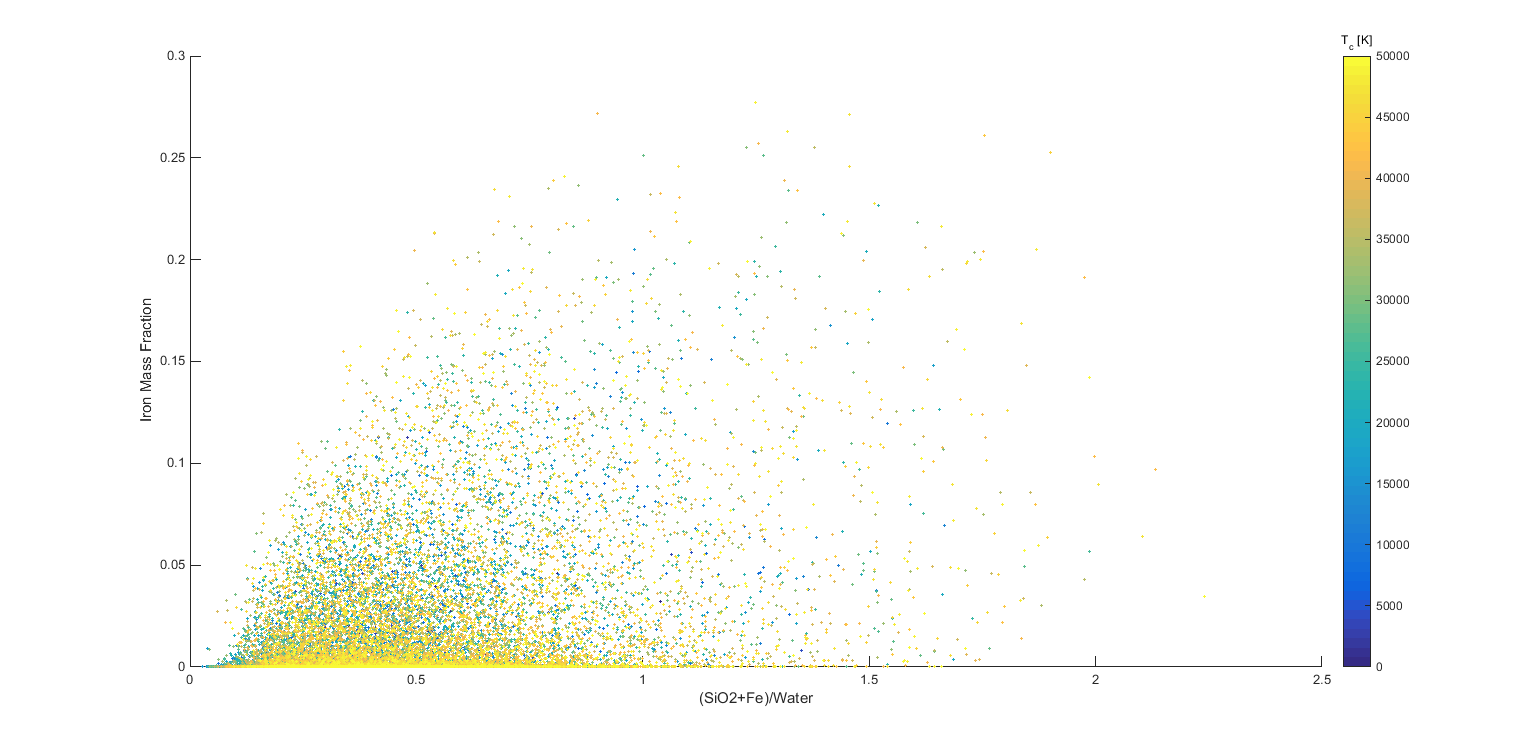}
	\caption{Mass fraction of iron (y-axis) vs. rock to water ratio (x-axis) for models computed using the inward algorithm. The color of the dots represents $T_c$ via the color bar on the right.  Shown are models with $T_c\leq 5\times 10^4$\,K.}
	\label{fig:fer}
\end{figure}

Fig.\,\ref{fig:er} shows the relation between the envelope mass fraction and the rock to water ratio for models computed with the inward algorithm with $T_c\leq 5\times 10^4$\,K.  One can see that as $T_c$ increases, there is a tendency for the rock to water ratio to increase as well at the lower envelope mass fractions.  This is due to the fact that higher central temperatures reduce the density of materials at any given pressure, so that the center of the planet can accommodate more \sio2 and iron.  In addition, as the rock to water ratio increases, the envelope mass fraction tends to increase as well.  This is necessary in order to keep the total density of the planet the same.  For the higher envelope mass fractions, however, the rock to water ratio seems to be independent of $T_c$, presumably because with sufficient low density envelope material, there is room for more high density rock.  Again, we see, in all cases, a hard upper limit of around 0.4 for the envelope mass fraction for all three $T_c$ regimes.  This is significantly higher than the values of $\sim 0.1-0.2$ found for classical models \citep{podcam74,podolak95,nettel12b,vazanhel2020}.  Some of these models have $T_c\leq 2.5\times 10^4$ and rock to water ratios of 1.2.  These are particularly interesting and will be studied in more detail in future work.

Fig.\,\ref{fig:fer} shows how the Fe mass fraction of the models varies with the rock to water ratio for inward algorithm models with $T_c\leq 5\times 10^4$\,K.  The largest Fe mass fractions seem to be associated with the higher rock to water ratios, but there are too few models at the highest rock to water ratios to be sure how far this trend continues.  The Fe mass fraction also seems to be independent of $T_c$ in this range.

	\begin{figure}[h!]
			\centering
			\includegraphics[height=6cm]{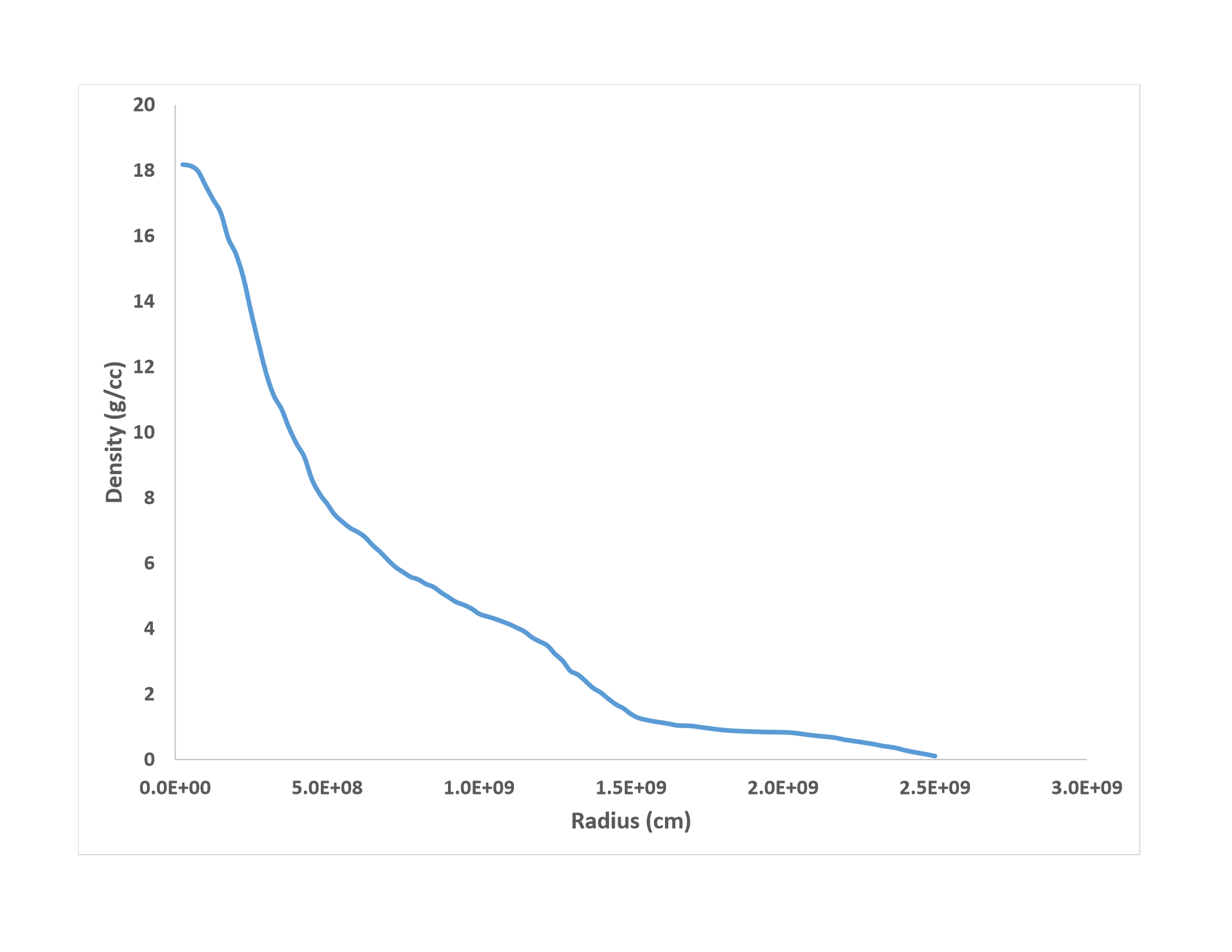}
			\includegraphics[height=6cm]{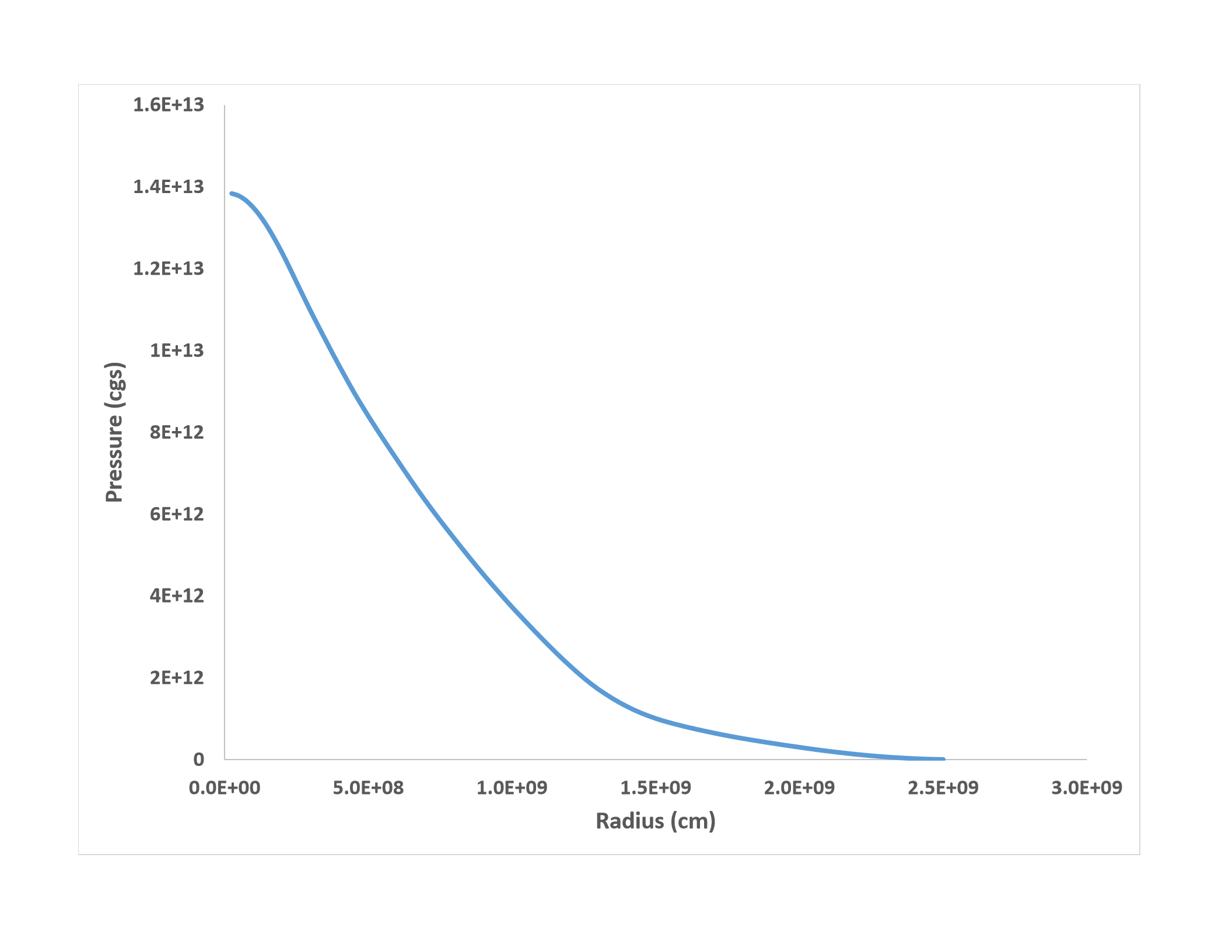}
			\includegraphics[height=6cm]{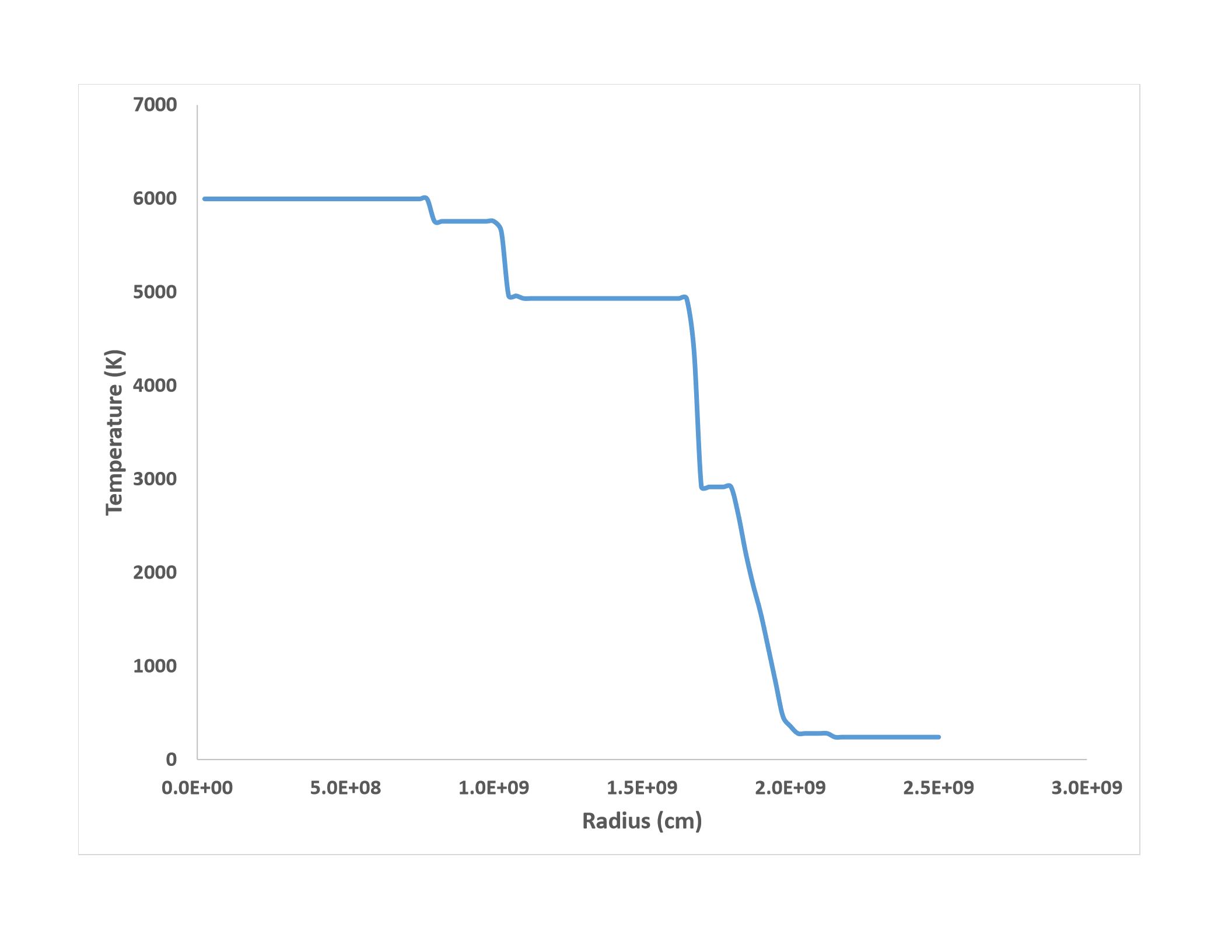}
			\includegraphics[height=6cm]{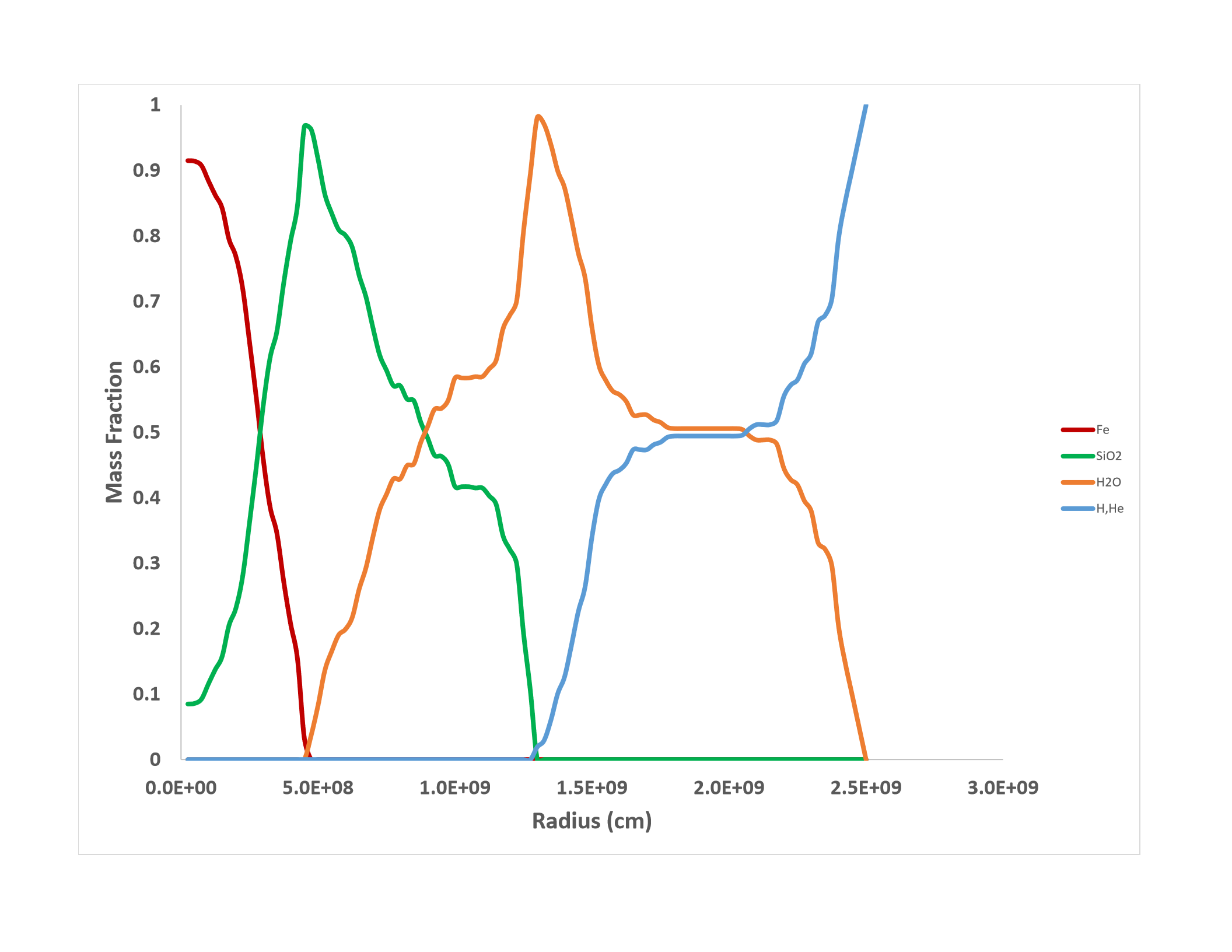}
			\caption{Structure of a sample model with a relatively low central temperature of 6000\,K.  The profiles of different parameters are shown as a function of radius.  {\bf Upper left:} Density. {\bf Upper right:} Pressure.  {\bf Lower left:} Temperature.  {\bf Lower right:} Mass fraction of iron (red), \sio2 (green), water (orange), and envelope (blue).}
			\label{fig:cmodel}
		\end{figure}
Fig.\,\ref{fig:cmodel} shows the density, pressure, temperature, and composition as a function of radius for a sample model that was chosen because it had a central temperature that was similar to that found in classical models, $T_c=5996$\,K.  As can be seen from the figure, this model has a smooth density distribution, and does not display the classical three-layer structure.  This particular model has 3.8\,$M_{\oplus}$ of rock, 7.5\,$M_{\oplus}$ of water, and 3.2\,$M_{\oplus}$ of envelope.  This gives a rock to water ratio of 0.5 which is very close to the solar value.  In short, from the standpoint of composition, this model is very similar to classical models of Uranus.  However, the temperature distribution within is far from adiabatic.  In particular there is a sharp drop in temperature of approximately 4700\,K between the radii of 16,300\,km and 20,300\,km.  Whether this is consistent with the assumed composition is something that needs to be investigated.  In principle, if such extreme temperature changes are shown to be unphysical, they can be identified and the resultant models automatically discarded.  Alternatively, the temperature profile can be numerically smoothed and the resultant composition recomputed.  
\begin{figure}[h!]
	\centering
	\includegraphics[height=6cm]{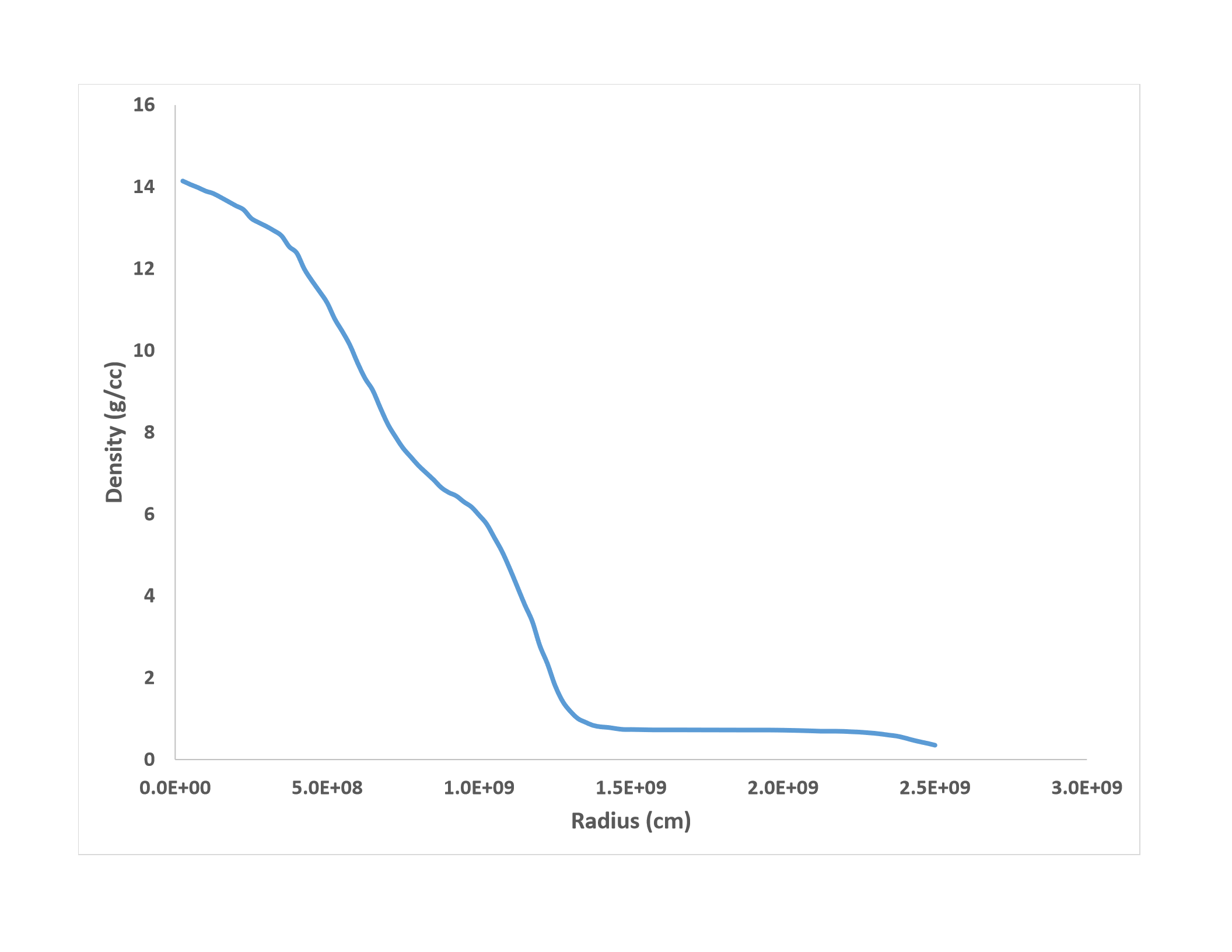}
	\includegraphics[height=6cm]{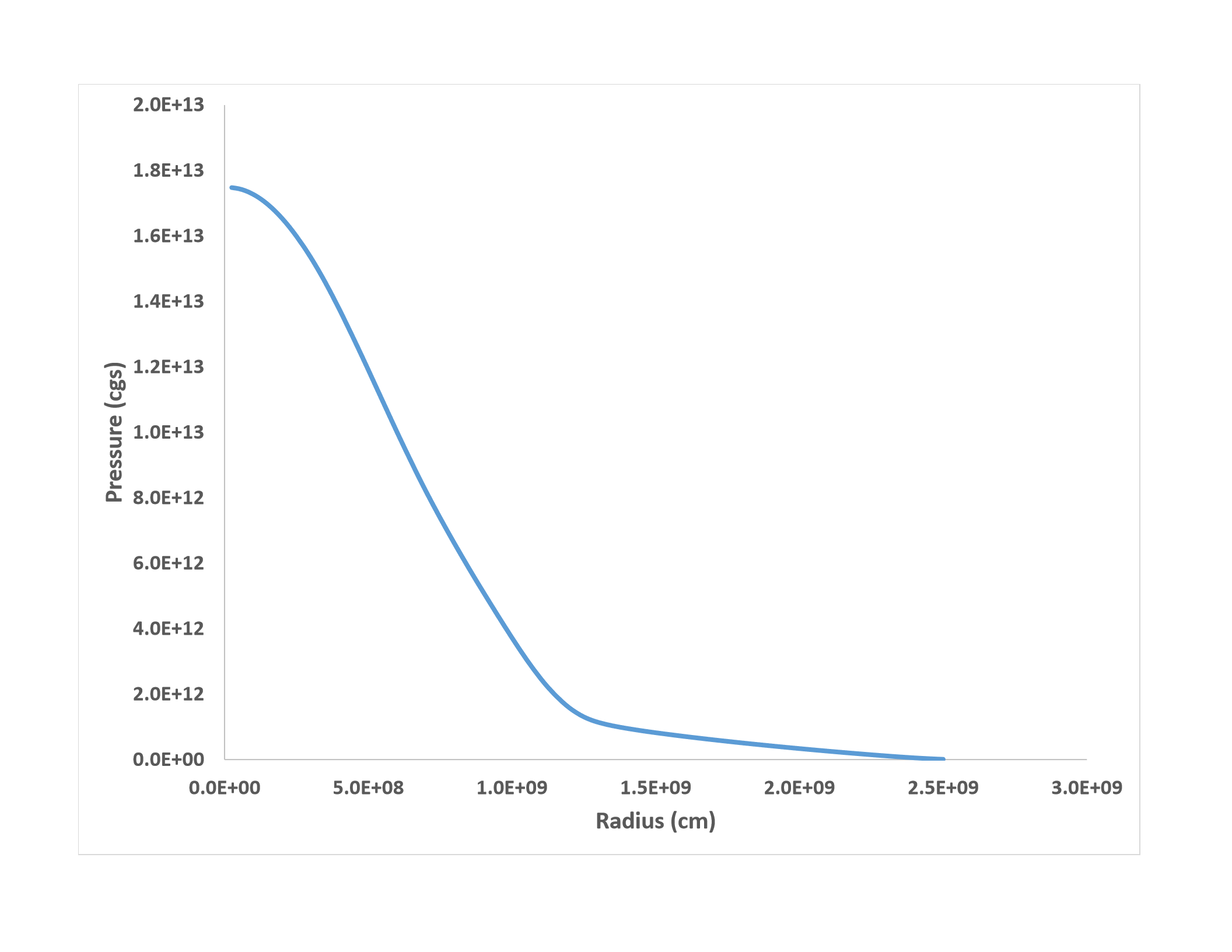}
	\includegraphics[height=6cm]{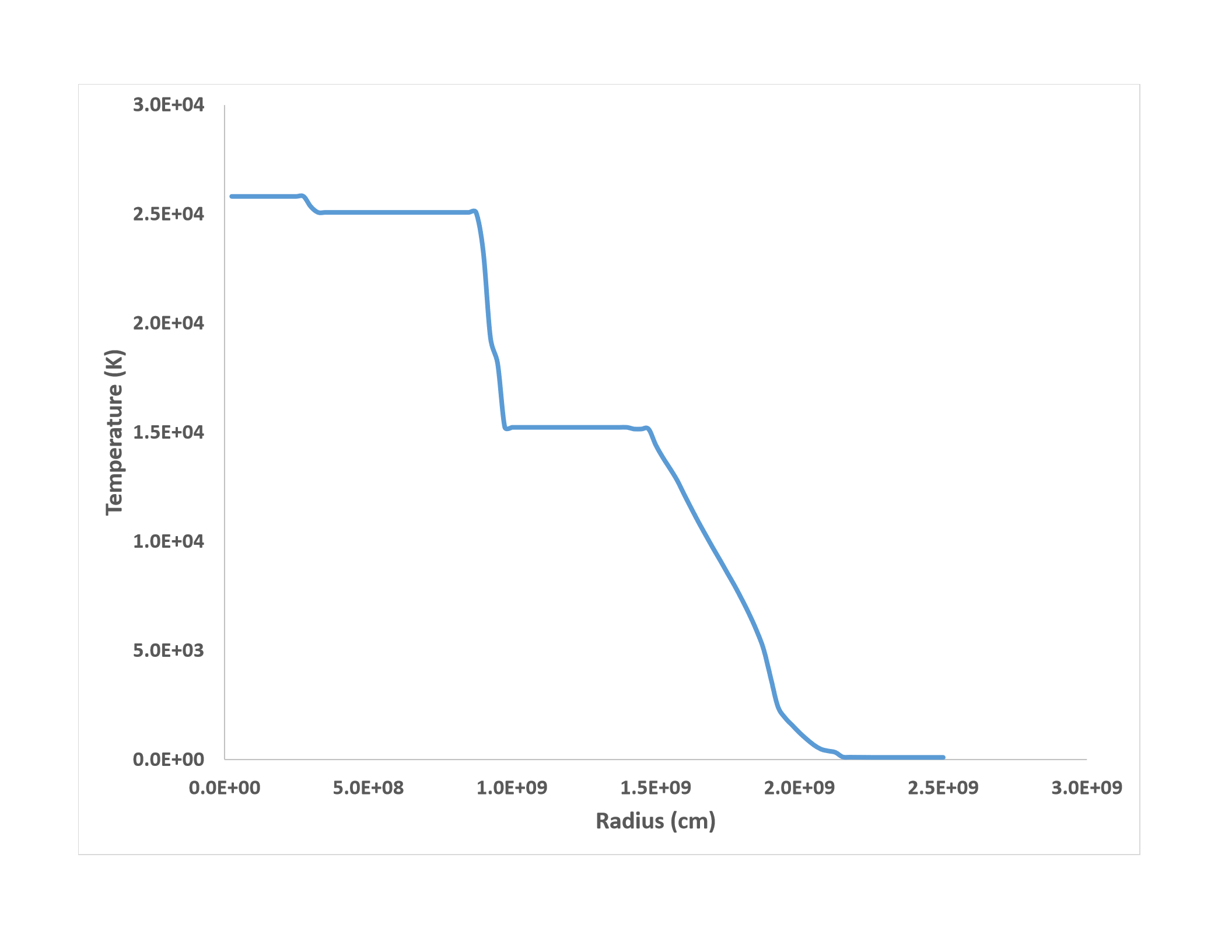}
	\includegraphics[height=6cm]{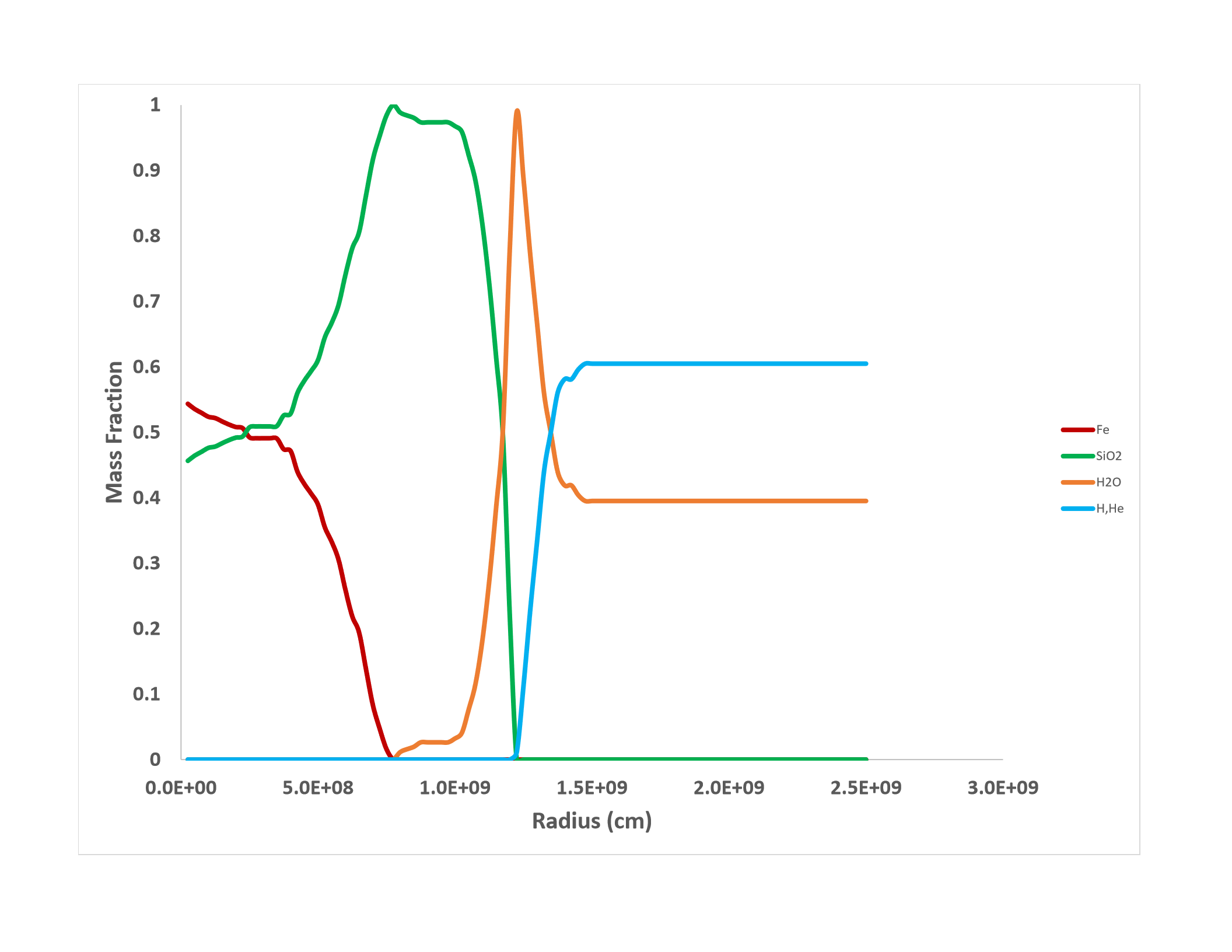}
	\caption{Same as Fig.\,\ref{fig:cmodel} but for a model with a rock to water ratio of 2.}
	\label{fig:hmodel}
\end{figure}

Fig.\,\ref{fig:hmodel} shows the structure of a model that has a rock to water ratio of 2.  The model consists of 7.1\,$M_{\oplus}$ of rock, 3.6\,$M_{\oplus}$ of water and 3.8\,$M_{\oplus}$ of envelope.  The central temperature is high, $2.58\times 10^4$\,K, but such temperatures are not out of the question \citep{podolak2019}.  A careful analysis of the model will need to be performed to determine whether the proposed temperature, pressure, density, and composition profiles are physically self-consistent, but this is beyond the scope of the present work.

\section{Summary}\label{S:Summary}
We have presented a method for generating random, monotonic, density distributions that are consistent with Uranus' mass, radius, and MOI.  Integrating the equations of mass conservation and hydrostatic equilibrium gives us a corresponding pressure profile.  We have also investigated three algorithms that can associate this density-pressure profile with temperature and composition profiles.  We have generated a suite of 100,000 such models in order to investigate the possible variation in the rock to water ratio in Uranus.  We find that although the great majority of models have ratios of 1 or less, it is possible to find models where the rock to water ratio is as high as 2 without having to mix envelope material into the core.  This high rock to water ratio is consistent with the observed composition of the small bodies in that region which are, presumably, representative of the building blocks of Uranus and Neptune (see references in Section\,\ref{S:Intro}).  The higher rock to water ratios require high central temperatures.  $T_c\sim 2.5\times 10^4$\,K for ratios of 2 or more but a ratio of 1.5 can be obtained with $T_c$ less than half that value, and this is within the range of central temperatures computed for classical models.  In addition, we find that there is an upper limit to the envelope mass fraction of 0.4.  This is higher, by roughly a factor of two, than the mass fractions generally cited, and opens the possibility that Uranus may contain much more hydrogen than is commonly supposed.

The temperature profiles we generate tend to have extended regions where the temperature is nearly constant, and places where the temperature drops abruptly.  This is the direct result of the algorithms we use, and needs to be checked for physical consistency.  These profiles can almost certainly be smoothed by imposing some small temperature gradient to the flat regions.  This should have very little effect on the associated composition profile.  We reiterate that these temperature-composition profiles that we have generated  need to be investigated in more detail to determine whether they are physically self-consistent, but we feel that they serve an important role in generating ``out of the box" thinking that could help us to understand, not only Uranus, but also the many different exoplanets that have been discovered. We see this work as a ``proof of concept", and future work will refine the criteria for generating temperature and composition profiles, and extend these models to other bodies in the solar system and beyond.

\section*{Acknowledgments}
\noindent M.P. was supported by a grant from the Pazy Fund of the Israel Atomic Energy Commission.

\bibliographystyle{icarus2}
\bibliography{paper} 
	
\newpage

\end{document}